\documentclass[prd,nofootinbib,twocolumn,showpacs,superscriptaddress]{revtex4}
\usepackage{graphics,rotating,multirow,bm,color,hyperref,amsmath,amssymb}
\hypersetup{
    colorlinks=true,
    linkcolor=green,
    citecolor=blue,
}

\begin{document}

\title{$N$-Body Simulations for Coupled Scalar Field Cosmology}
\author{Baojiu~Li}
\email[Email address: ]{b.li@damtp.cam.ac.uk}
\affiliation{DAMTP, Centre for Mathematical Sciences, University of Cambridge,
Wilberforce Road, Cambridge CB3 0WA, UK}
\affiliation{Kavli Institute for Cosmology Cambridge, Madingley Road, Cambridge CB3 0HA,
UK}
\author{John~D.~Barrow}
\email[Email address: ]{j.d.barrow@damtp.cam.ac.uk}
\affiliation{DAMTP, Centre for Mathematical Sciences, University of Cambridge,
Wilberforce Road, Cambridge CB3 0WA, UK}
\date{\today}

\begin{abstract}
We describe in detail the general methodology and numerical implementation
of consistent N-body simulations for coupled scalar field cosmological
models, including the background cosmology and the generation of initial
conditions (with the different couplings to different matter species taken
into account). We perform fully consistent simulations for a class of
coupled scalar field models with an inverse power-law potential and negative
coupling constant, for which the chameleon mechanism does not operate. We
find that in such cosmological models the scalar-field potential plays a
negligible role except in the background expansion, and the fifth force that
is produced is proportional to gravity in magnitude, justifying the use of a
rescaled gravitational constant G in some earlier N-body simulations of
similar models. We study the effects of the scalar coupling on the nonlinear
matter power spectra and compare with linear perturbation calculations to
investigate where the nonlinear model deviates from the linear
approximation. For the first time, the algorithm to identify gravitationally
virialized matter halos is adapted to the scalar field cosmology, and then
used to measure the mass function and study the properties of virialized
halos. We find that the net effect of the scalar coupling helps produce more
heavy halos in our simulation boxes and suppresses the inner (but not the
outer) density profile of halos compared with those predicted by lambda-CDM,
while this suppression weakens as the coupling between the scalar field and
dark matter particles increases in strength.

\end{abstract}

\maketitle

\section{Introduction}

\label{sect:intro}

The nature of the dark energy \cite{Copeland:2006} driving an apparent
acceleration of the universe has been a cosmological puzzle for more than a
decade. Amongst the models proposed to explain it, those incorporating
scalar fields are by far the most popular, not only because of their
mathematical simplicity and phenomenological richness, but also because the
scalar field is a natural ingredient of many high-energy physics theories. A
scalar field contributes a single dynamical degree of freedom which can
interact indirectly with other matter species through gravity or couple
directly to matter, producing a fifth force on the matter which creates
violations of the Weak equivalence principle (WEP). This second possibility
was introduced with the hope that such a coupling could potentially
alleviate the coincidence problem of dark energy \cite{Amendola:2000} and
has since then attracted much attention (see, for example,\emph{~}\cite%
{Bean:2001, Amendola:2004, Koivisto:2005, Boehmer:2008, Bean:2008,
Bean:2008b, Boehmer:2009} and references therein for some recent work).

If there is a direct coupling between the scalar field and baryons, then the
baryonic particles will experience a fifth force, which is severely
constrained by observations, unless there is some special mechanism
suppressing the fifth-force effects. This is the case in chameleon models,
where the scalar field (the chameleon) gains mass in high-density regions
(where observations and experiments are performed) and the fifth force
effects are confined to undetectably small distances \cite{Khoury:2004a,
Khoury:2004b, Mota:2006, Mota:2007}. A common approach which avoids such
complications is to assume that the scalar field couples only to the dark
matter, an idea seen frequently in models with a coupled dark sector (\emph{%
e.g.}~\cite{Maartens:2009, Valiviita:2010, Peacock:2010}). In this work our
scalar field will not be chameleon-like as this case has been investigated
elsewhere \cite{Li:2009sy, Zhao:2010, Li:2010}.

The formation of cosmological structure in the presence of coupled scalar
fields at the linear perturbative level has already been studied in great
detail. Here, we extend these studies into the nonlinear regime. In
particular, we want to know how structure formation on the scales of
galaxies and galaxy clusters is modified. As we shall see below, there are
principally four effects of the scalar field coupling, namely through (i)
the modification of the background expansion rate, (ii) the action of a
fifth force on dark-matter particles, (iii) the reduced contribution of dark
matter to the Poisson equation, and (iv) the change of the initial condition
at early times. Of these four, (ii) can actually be subdivided into (a) an
essential "rescaling" of gravitational constant and (b) a velocity-dependent
acceleration term, while (iv) is a combined consequence of (i, ii, iii)
before $z\sim 50$. We cannot track all these effects into the nonlinear
regime using linear analysis and the tool we will employ to obtain
quantitative predictions is $N$-body simulation \cite{Bertschinger:1998}.

There has been some earlier work in this area (see for example,\emph{~}\cite%
{Linder:2003, Mainini:2003, Springel:2007, Kesden:2006, Farrar:2007,
Keselman:2009, Maccio:2004, Baldi:2008, Rodriguez-Meza:2007,
Rodriguez-Meza:2008, Rodriguez-Meza:2009, Hellwing:2009} and also \cite%
{Oyaizu:2008a, Oyaizu:2008b, Laszlo:2008, Schmidt:2009, Chan:2009,
Zhang:2010} for related works). However, in all those scalar-field
simulations the scalar field equation is not solved explicitly, but rather
it is assumed either that the fifth force is proportional to gravity so that
its effect is a simple rescaling of the gravitational constant, or that the
fifth force takes the Yukawa form with exponential cut-off beyond some
scale. On the other hand, the recent fully consistent simulations performed
in \cite{Li:2009sy, Zhao:2010, Li:2010} have shown that, at least for the
chameleon scalar field models, the above simplifying approximations are not
good and raises questions about the extent to which we should trust them. It
is these concerns that motivates this paper, in which we test the accuracy
of those approximations, and study in more detail the qualitative and
quantitative effects of a coupled scalar field on the formation and
evolution of the nonlinear cosmic structure.

As in \cite{Li:2010, Maccio:2004}, we shall consider a universe which at
late times is dominated by a scalar field and two matter species, namely the
dark matter, which couples to the scalar field in a way prescribed in Sect.~%
\ref{subsect:basiceqn} and the so-called "baryons", which are essentially
the dark matter \emph{without} scalar-field coupling.

The paper is organized as follows: in Sect~\ref{sect:eqn} we list the
essential equations to be implemented by our $N$-body simulations and
describe briefly their differences from standard $\Lambda $CDM. Sect.~\ref%
{sect:simu} presents a comprehensive description of the methodology used in
this paper and its implementation in the numerical code. Sect.~\ref%
{subsect:code} introduces the code, Sect.~\ref{subsect:param} lists the
physical and simulation parameters we adopt, Sect.~\ref{subsect:baryons_inc}
illustrates how we distinguish between baryons and dark matter particles,
Sect.~\ref{subsect:bkgd} summarizes the results for the background expansion
and linear perturbation evolution, which will be referred to subsequently
from time to time, and finally Sect.~\ref{subsect:ic} and Appendix~\ref%
{appen:Zeldovich} explain in detail how to generate initial conditions for
the $N$-body simulations which take into account the effects of the coupling
between dark matter and the scalar field. As we aim to set up a general
framework for $N$-body simulations of coupled scalar field models, we have
tried to include all the main ingredients in this section. Our numerical
results are presented in Sect.~\ref{sect:results}, within which Sect.~\ref%
{subsect:snapshots} displays some general results, showing that the
approximation of rescaling gravitational constant as used in previous
literature is a very good one for the model studied here (but not
necessarily so for other models!), and Sects.~\ref{subsect:Pk}, \ref%
{subsect:mf} and \ref{subsect:halo} discuss, respectively, how the coupling
modifies the nonlinear matter power spectrum, mass function and the internal
density profiles of halos. We present our conclusions in \ref{sect:con}.

All through the paper a subscript $_{\mathrm{B}}$ ($_{\mathrm{D}}$) denotes
a corresponding quantity for baryons (dark matter), unless otherwise stated.

\section{The Equations}

The equations that will be used in the $N$-body simulations have
been discussed in detail in Ref.~\cite{Li:2009sy, Li:2010}. As we
are considering a qualitatively different model here, and the
parameterization and discretization of equations are both
different, we list these equations and discuss them for
completeness.

\label{sect:eqn}

\subsection{The Basic Equations}

\label{subsect:basiceqn}

The Lagrangian for our coupled scalar field model is
\begin{equation}\label{eq:Lagrangian}
\mathcal{L}=\frac{1}{2}\left[ \frac{R}{\kappa }-\nabla ^{a}\varphi
\nabla
_{a}\varphi \right] +V(\varphi )-C(\varphi )\mathcal{L}_{\mathrm{CDM}}+%
\mathcal{L}_{\mathrm{S}}\ ,\
\end{equation}%
where $R$ is the Ricci scalar, $\kappa =8\pi G,$ with $G$ Newton's constant,
$\varphi $ is the scalar field, $V(\varphi )$ its potential energy, and $%
C(\varphi )$ its coupling to the dark matter, which is assumed to be cold
and described by the Lagrangian $\mathcal{L}_{\mathrm{CDM}}$; $\mathcal{L}_{%
\mathrm{S}}$ includes all other matter species, including the \emph{baryons}%
. The contributions from photons and neutrinos in the $N$-body simulations
(for late times, $z\sim \mathcal{O}(1)$) is negligible, but should be
included when generating the matter power-spectrum at redshift $z\sim
\mathcal{O}(50)$, which depends on the early evolution of the universe, from
which the initial conditions for our $N$-body simulations are obtained (see
Sect.~\ref{subsect:ic} and Appendix~\ref{appen:Zeldovich}).

The dark-matter Lagrangian for a point particle with (bare) mass $m_{0}$ is
\begin{equation}\label{eq:DMLagrangian}
\mathcal{L}_{\mathrm{CDM}}(\mathbf{y})=-\frac{m_{0}}{\sqrt{-g}}\delta (%
\mathbf{y}-\mathbf{x}_{0})\sqrt{g_{ab}\dot{x}_{0}^{a}\dot{x}_{0}^{b}},
\end{equation}%
where $\mathbf{y}$ is the general coordinate and $\mathbf{x}_{0}$ is the
coordinate of the centre of the particle. From this equation we derive the
corresponding energy-momentum tensor:
\begin{equation}\label{eq:DMEMT_particle}
T_{\mathrm{CDM}}^{ab}=\frac{m_{0}}{\sqrt{-g}}\delta (\mathbf{y}-\mathbf{x}%
_{0})\dot{x}_{0}^{a}\dot{x}_{0}^{b}.
\end{equation}%
Also, because $g_{ab}\dot{x}_{0}^{a}\dot{x}_{0}^{b}\equiv
g_{ab}u^{a}u^{b}=1, $ where $u^{a}$ is the four-velocity of the dark-matter
particle centred at $x_{0}$, the Lagrangian can be rewritten as
\begin{equation}\label{eq:DMLagrangian2}
\mathcal{L}_{\mathrm{CDM}}(\mathbf{y})=-\frac{m_{0}}{\sqrt{-g}}\delta (%
\mathbf{y}-\mathbf{x}_{0}),
\end{equation}%
which will be used below (see Appendix~\ref{appen:Lagrangian}).

Eq.~(\ref{eq:DMEMT_particle}) is just the energy-momentum tensor for a
single dark matter particle. For a fluid with many particles, the
energy-momentum tensor will be
\begin{eqnarray}\label{eq:DMEMT_fluid}
T_{\mathrm{CDM}}^{ab} &=&\frac{1}{\mathcal{V}}\int_{\mathcal{V}}d^{4}y\sqrt{%
-g}\frac{m_{0}}{\sqrt{-g}}\delta (y-x_{0})\dot{x}_{0}^{a}\dot{x}_{0}^{b}
\notag\\
&=&\rho_{\mathrm{CDM}}u^{a}u^{b},
\end{eqnarray}%
in which $\mathcal{V}$ is a volume that is microscopically large but
macroscopically small, and we have extended the 3-dimensional $\delta $
function to a 4-dimensional one by adding a time component. Here, $u^{a}$ is
the averaged four-velocity of the dark-matter fluid, which is \emph{not}
necessarily the same as the four-velocity of the observer, a point which we
discuss below.

Using
\begin{equation}
T^{ab}=-\frac{2}{\sqrt{-g}}\frac{\delta \left( \sqrt{-g}\mathcal{L}\right) }{%
\delta g_{ab}},
\end{equation}%
it is straightforward to show that the energy-momentum tensor for the scalar
field is
\begin{equation}\label{eq:phiEMT}
T^{\varphi ab}=\nabla ^{a}\varphi \nabla ^{b}\varphi -g^{ab}\left[ \frac{1}{2%
}\nabla _{c}\varphi \nabla ^{c}\varphi -V(\varphi )\right] .
\end{equation}%
Therefore the total energy-momentum tensor is
\begin{eqnarray}\label{eq:EMT_tot}
T_{ab} &=&\nabla _{a}\varphi \nabla _{b}\varphi -g_{ab}\left[ \frac{1}{2}%
\nabla _{c}\varphi \nabla ^{c}\varphi -V(\varphi )\right]  \notag
\\
&&+C(\varphi )T_{ab}^{\mathrm{CDM}}+T_{ab}^{\mathrm{S}}
\end{eqnarray}%
where $T_{ab}^{\mathrm{CDM}}=\rho _{\mathrm{CDM}}u_{a}u_{b}$, $T_{ab}^{%
\mathrm{S}}$ is the energy-momentum tensor for all other matter species
including baryons, and the Einstein equations are
\begin{equation}\label{eq:EinsteinEq}
G_{ab}=\kappa T_{ab}
\end{equation}%
where $G_{ab}$ is the Einstein tensor. Note that because of the extra
coupling between the scalar field, $\varphi $, and the dark matter, the
energy-momentum tensors for either will not be separately conserved, and we
have
\begin{equation}\label{eq:DM_energy_conservation}
\nabla _{b}T^{\mathrm{CDM}_{{}}ab}=-\frac{C_{\varphi }(\varphi )}{C(\varphi )%
}\left( g^{ab}\mathcal{L}_{\mathrm{CDM}}+T^{\mathrm{CDM}ab}\right) \nabla
_{b}\varphi ,\ \
\end{equation}%
where throughout this paper we shall use subscript $_{\varphi }$ to denote a
derivative with respect to $\varphi $. However, the total energy-momentum
tensor is conserved.

Finally, the scalar field equation of motion (EOM) is
\begin{equation}
\square \varphi +\frac{\partial V(\varphi )}{\partial \varphi }=\frac{%
\partial C(\varphi )}{\partial \varphi }\mathcal{L}_{\mathrm{CDM}},
\end{equation}%
where $\square \equiv \nabla ^{a}\nabla _{a}$. Using Eq.~(\ref%
{eq:DMLagrangian2}), it can be rewritten as
\begin{equation}\label{eq:phiEOM}
\square \varphi +\frac{\partial V(\varphi )}{\partial \varphi }+\rho _{%
\mathrm{CDM}}\frac{\partial C(\varphi )}{\partial \varphi }=0.
\end{equation}

Eqs.~(\ref{eq:EMT_tot}, \ref{eq:EinsteinEq}, \ref{eq:DM_energy_conservation}%
, \ref{eq:phiEOM}) summarize all the physics needed for the following
analysis.

We will consider an inverse power-law potential energy for the scalar field,
\begin{equation}\label{eq:potential}
V(\varphi )=\frac{\Lambda ^{4}}{\left( \sqrt{\kappa }\varphi
\right) ^{\alpha }},
\end{equation}
where $\alpha $ is a dimensionless constant and $\Lambda $ is a
constant with dimensions of mass. This potential has also been
adopted in various background or linear perturbation studies of
scalar fields (either minimally or non-minimally coupled); the
tracking behaviour its produces makes it a good dark energy
candidate and for that purpose we shall choose $\alpha \sim
\mathcal{O}(0.1-1)$. Meanwhile, the coupling between the scalar
field and dark matter particles is chosen as
\begin{equation}\label{eq:coupling_function}
C(\varphi )=\exp (\gamma \sqrt{\kappa }\varphi ),
\end{equation}%
where $\gamma <0$ is another dimensionless constant characterizing the
strength of the coupling. As we shall see below, $2|\gamma |^{2}$ is roughly
the ratio of the magnitudes of the fifth force and gravity on the dark
matter particles.

The bare potential Eq.~(\ref{eq:potential}) and the coupling function Eq.~(%
\ref{eq:coupling_function}) form an effective total potential
\begin{equation}
V_{eff}(\varphi )=V(\varphi )+\rho _{\mathrm{CDM}}C(\varphi )
\end{equation}%
for the scalar field $\varphi $, just as in \cite{Li:2009sy, Li:2010}.
However, both $V(\varphi )$ and $C(\varphi )$ decrease as $\varphi $
increases here, there is no finite global minimum for $V_{eff}(\varphi ),$
and the scalar field will always continue rolling down the potential given
appropriate initial condition. As a result, although in \cite{Li:2009sy,
Li:2010} the scalar field almost always resides around the minimum of $%
V_{eff}(\varphi )$, where it acquires a heavy mass to become a chameleon
\cite{Khoury:2004a, Khoury:2004b, Mota:2006, Mota:2007}, this does not
necessarily happen here. Instead, the rolling of the scalar field can be
quite rapid, introducing interesting new dynamics in both background
cosmology and perturbation evolution.

\subsection{The Non-Relativistic Limits}

\label{subsect:nonrel}

The $N$-body simulation only probes the motion of particles at late times,
and we are not interested in extreme conditions such as black hole formation
and evolution, so that we can take the non-relativistic limit of the above
equations as a good approximation.

As discussed in \cite{Li:2010}, the existence of the scalar field and its
couplings to matter particles leads to several changes compared with the $%
\Lambda $CDM paradigm:

\begin{enumerate}
\item The scalar field has its own energy-momentum tensor, which could
change the source term of the Poisson equation because the scalar field,
unlike the cosmological constant, can cluster.

\item The mass of the dark matter particles is \emph{effectively}
renormalized because of the coupling to the scalar field (we say
"effectively" because the bare physical mass itself is unchanged but it is
multiplied by the coupling function $C(\varphi )$ when it appears in
gravitational equations and the scalar field equation of motion).

\item The scalar field contributes a fifth force on the dark matter
particles, so they no longer follow geodesics determined by gravity only.

\item As the dark matter and baryonic particles couple to the scalar field
with different strengths, they will influence the scalar field in different
ways, feel different fifth forces, and their subsequent motions will differ.
\end{enumerate}

It therefore becomes clear that the following equations, or quantities, in
their non-relativistic forms, are needed:

\begin{enumerate}
\item The scalar field equation of motion, to compute the value of the
scalar field $\varphi $ at any given time and position;

\item The Poisson equation, to determine the gravitational potential at any
given time and position from the local energy density and pressure, which
includes the contribution from the scalar field (obtained from $\varphi $
equation of motion);

\item The total force on the dark matter particles, determined by the
spatial configuration of $\varphi $, and like gravity it is determined by
the spatial configuration of the gravitational potential;

\item The total force on baryonic particles, which receives no contribution
from the scalar field $\varphi $. It is solely determined by the spatial
configuration of the gravitational potential.
\end{enumerate}

We shall describe these in turn. For the scalar field equation of motion, we
denote by $\bar{\varphi}$ the background value of $\varphi $ and $\delta
\varphi \equiv \varphi -\bar{\varphi}$. Then Eq.~(\ref{eq:phiEOM}) can be
rewritten as
\begin{eqnarray}
\ddot{\delta\varphi} + 3H\dot{\delta\varphi} +
\vec{\nabla}_{\mathbf{r}}^{2}\varphi +
V_{,\varphi}(\varphi)-V_{,\varphi}(\bar{\varphi})\nonumber\\ +
\rho_{\mathrm{CDM}}C_{,\varphi}(\varphi) -
\bar{\rho}_{\mathrm{CDM}}C_{,\varphi}(\bar{\varphi}) &=&
0\nonumber
\end{eqnarray}
by removing the background part. Here,
$\vec{\nabla}_{\mathbf{r}a}$ is the
covariant spatial derivative with respect to the physical coordinate $%
\mathbf{r}=a\mathbf{x,}$ $\mathbf{x}$ is the comoving coordinate, and $\vec{%
\nabla}_{\mathbf{r}}^{2}=\vec{\nabla}_{\mathbf{r}a}\vec{\nabla}_{\mathbf{r}%
}^{a}$. $\vec{\nabla}_{\mathbf{r}a}$ is strictly speaking non-Euclidian as
the spacetime is not completely flat, but because we are working in the weak
field limit we approximate it as Euclidian, that is $\vec{\nabla}_{\mathbf{r}%
}^{2}\doteq -\left( \partial _{r_{x}}^{2}+\partial _{r_{y}}^{2}+\partial
_{r_{z}}^{2}\right) $; the minus sign is because our metric convention is $%
(+,-,-,-)$.

In our simulations we also work in the quasi-static limit, and assume that
the spatial gradients are much larger than the time derivatives, $|\vec{%
\nabla}_{\mathbf{r}}\varphi |\gg |\frac{\partial \delta \varphi }{\partial t}%
|$. Therefore, the above equation is simplified to
\begin{eqnarray}\label{eq:WFphiEOM}
&&c^{2}\partial_{\mathbf{x}}^{2}(a\delta\varphi)\\ &=&
a^{3}\left[V_{,\varphi}(\varphi)-V_{,\varphi}(\bar{\varphi}) +
\rho_{\mathrm{CDM}}C_{,\varphi}(\varphi) -
\bar{\rho}_{\mathrm{CDM}}C_{,\varphi}(\bar{\varphi})\right],\nonumber
\end{eqnarray}
in which $\partial
_{\mathbf{x}}^{2}=-\vec{\nabla}_{\mathbf{x}}^{2}=+\left(
\partial _{x}^{2}+\partial _{y}^{2}+\partial _{z}^{2}\right) $ is with
respect to $\mathbf{x,}$ with $\vec{\nabla}_{\mathbf{x}}=a\vec{\nabla}_{%
\mathbf{r}}$, and we have restored the factor $c^{2}$ in front of $\vec{%
\nabla}_{\mathbf{x}}^{2}$ (the $\varphi $ here and in the remaining of this
paper is $c^{-2}$ times the $\varphi $ in the original Lagrangian unless
otherwise stated). Note that here $V$ and $\rho _{\mathrm{CDM}}$ both have
the dimensions of\emph{\ mass }density rather than \emph{energy} density.

Next consider the Poisson equation, which is obtained from the Einstein
equation in the weak-field and slow-motion limits. Here the metric can be
written as
\begin{eqnarray}
ds^{2} &=& (1+2\phi)dt^{2} - (1-2\psi)\delta_{ij}dr^{i}dr^{j}
\end{eqnarray}
from which we find that the time-time component of the Ricci
curvature tensor $R_{\ 0}^{0}=-\vec{\nabla}_{\mathbf{r}}^{2}\phi $
. The Einstein equation gives
\begin{eqnarray}\label{eq:EinsteinEqn}
R^{0}_{\ 0}\ =\ -\vec{\nabla}_{\mathbf{r}}^{2}\phi\ =\
\frac{\kappa}{2}(\rho_{\mathrm{TOT}}+3p_{\mathrm{TOT}})
\end{eqnarray}
where $\rho _{\mathrm{TOT}}$ and $p_{\mathrm{TOT}}$ are the total
energy
density and pressure, respectively. The quantity $\vec{\nabla}_{\mathbf{r}%
}^{2}\phi $ can be expressed in terms of the comoving coordinate $\mathbf{x}$
as
\begin{eqnarray}
\vec{\nabla}_{\mathbf{r}}^{2}\phi &=& \frac{1}{a^{2}}
\vec{\nabla}_{\mathbf{x}}^{2}\left(\frac{\Phi}{a} -
\frac{1}{2}a\ddot{a}\mathbf{x}^{2}\right)\nonumber\\
&=& \frac{1}{a^{3}}\vec{\nabla}_{\mathbf{x}}^{2}\Phi -
3\frac{\ddot{a}}{a}
\end{eqnarray}
where we have defined a new Newtonian potential
\begin{eqnarray}\label{eq:newphi}
\Phi &\equiv& a\phi + \frac{1}{2}a^{2}\ddot{a}\mathbf{x}^{2}
\end{eqnarray}
Thus,
\begin{eqnarray}
\vec{\nabla}_{\mathbf{x}}^{2}\Phi &=&
a^{3}\left(\vec{\nabla}_{\mathbf{r}}^{2}\phi +
3\frac{\ddot{a}}{a}\right)\\
&=& -
a^{3}\left[\frac{\kappa}{2}(\rho_{\mathrm{TOT}}+3p_{\mathrm{TOT}})
-
\frac{\kappa}{2}(\bar{\rho}_{\mathrm{TOT}}+3\bar{p}_{\mathrm{TOT}})\right]\nonumber
\end{eqnarray}
where in the second step we have used Eq.~(\ref{eq:EinsteinEqn})
and the Raychaudhuri equation, and an overbar labels the
background value of a quantity. Because the energy-momentum tensor
for the scalar field is given by Eq.~(\ref{eq:phiEMT}), it is easy
to show that $\rho ^{\varphi }+3p^{\varphi }=2\left[
\dot{\varphi}^{2}-V(\varphi )\right] $ and so
\begin{eqnarray}
\vec{\nabla}_{\mathbf{x}}^{2}\Phi &=& -4\pi
Ga^{3}\left\{\rho_{\mathrm{CDM}}C(\varphi)+\rho_{\mathrm{B}} +
2\left[\dot{\varphi}^{2}-V(\varphi)\right]\right\}\nonumber\\
&& +4\pi Ga^{3} \left\{\bar{\rho}_{\mathrm{CDM}}C(\bar{\varphi})+
\bar{\rho}_{\mathrm{B}}+
2\left[\dot{\bar{\varphi}}^{2}-V(\bar{\varphi})\right]\right\}.\nonumber
\end{eqnarray}
In this equation $\dot{\varphi}^{2}-\dot{\bar{\varphi}}^{2}=2\dot{\varphi}%
\dot{\delta \varphi }+\dot{\delta \varphi }^{2}\ll (\vec{\nabla}_{\mathbf{r}%
}\varphi )^{2}$ in the quasi-static limit, and so could be dropped safely.
Therefore, we have
\begin{eqnarray}\label{eq:WFPoisson}
\partial_{\mathbf{x}}^{2}\Phi &=& 4\pi Ga^{3}
\left[\rho_{\mathrm{CDM}}C(\varphi)-\bar{\rho}_{\mathrm{CDM}}C(\bar{\varphi})\right]\nonumber\\
&&+4\pi Ga^{3}
\left[\rho_{\mathrm{B}}-\bar{\rho}_{\mathrm{B}}\right] - 8\pi
Ga^{3}\left[V(\varphi)-V(\bar{\varphi})\right].\ \
\end{eqnarray}

Finally, for the equations of motion of the dark matter particles, consider
Eq.~(\ref{eq:DM_energy_conservation}). Using Eqs.~(\ref{eq:DMEMT_particle}, %
\ref{eq:DMLagrangian2}), this can be reduced to
\begin{eqnarray}\label{eq:DMEOM}
\ddot{x}^{a}_{0} + \Gamma^{a}_{bc}\dot{x}^{b}_{0}\dot{x}^{c}_{0}
&=&
\left(g^{ab}-u^{a}u^{b}\right)\frac{C_{\varphi}(\varphi)}{C(\varphi)}\nabla_{b}\varphi.
\end{eqnarray}
In this equation, the left-hand side is the conventional geodesic
equation of general relativity, and the right-hand side is the new
fifth force
contribution from the coupling to the scalar field. Note that $%
g^{ab}-u^{a}u^{b}=h^{ab}$ is the projection tensor that projects any
4-tensor into the 3-space perpendicular to $u^{a}$, so $\left(
g^{ab}-u^{a}u^{b}\right) \nabla _{a}=\hat{\nabla}^{b}$ is the spatial
derivative in the 3-space of the observer and perpendicular to $u^{a}$.
Therefore, the right-hand side of Eq.~(\ref{eq:DMEOM}), $\frac{C_{\varphi
}(\varphi )}{C(\varphi )}\hat{\nabla}_{a}\varphi =\hat{\nabla}_{a}\log
C(\varphi )$, is perpendicular to $u^{a}$, which shows that the energy
density of dark matter is conserved and only the particle trajectory is
modified \cite{Li:2009sy}. As the 'observers' are just the individual dark
matter particles [cf.~Eq.~(\ref{eq:DMEMT_particle})], the force computed
above is exactly what those particles feel. The $\delta \varphi $ in Eq.~(%
\ref{eq:DMEOM}) is measured in the dark-matter rest frame while that in Eq.~(%
\ref{eq:WFphiEOM}) is measured in the fundamental observer's frame. Then, in
order to use the $\delta \varphi $ obtained from Eq.~(\ref{eq:WFphiEOM}) in
Eq.~(\ref{eq:DMEOM}), we need to perform a gauge transformation $\hat{\nabla}%
_{a}\varphi \rightarrow \hat{\nabla}_{a}\varphi -\dot{\varphi}v_{a}$ where $%
v_{a}$ is the peculiar velocity of the dark-matter particle relative to the
fundamental observer and consequently we will have an additional term $\frac{%
C_{\varphi }}{C}\dot{\varphi}v^{a}$ in Eq.~(\ref{eq:DMEOM}) which
is the velocity-dependent term identified in \cite{Maccio:2004}.
From here on we will always use the $\hat{\nabla}_{a}\varphi $
measured in the fundamental observer's frame (in which the density
perturbation is obtained more directly), and so
Eq.~(\ref{eq:DMEOM}) should be changed to
\begin{eqnarray}\label{eq:DMEOM1}
\ddot{x}^{a}_{0} + \Gamma^{a}_{bc}\dot{x}^{b}_{0}\dot{x}^{c}_{0}
&=&
\frac{C_{\varphi}(\varphi)}{C(\varphi)}\left(\hat{\nabla}^{a}\varphi-\dot{\varphi}v^a\right).
\end{eqnarray}
We stress that here
\begin{eqnarray}
v^{i}\ =\ \dot{r}^{i}\ =\ a\dot{x}^{i} + \dot{a}x^{i}\ =\
a\dot{x}^{i}
\end{eqnarray}
where $\dot{r}^{i}$ is the velocity of a dark-matter particle
relative to the fundamental observer which is at the same position
at that moment so that $x^{i}=0$ but $\dot{x}^{i}\neq 0$. In the
non-relativistic limit, the spatial components of
Eq.~(\ref{eq:DMEOM1}) can be written as
\begin{eqnarray}\label{eq:DMEOM_physical}
\frac{d^{2}\mathbf{r}}{dt^{2}} &=& -\vec{\nabla}_{\mathbf{r}}\phi
-
\frac{C_{\varphi}(\varphi)}{C(\varphi)}\vec{\nabla}_{\mathbf{r}}\varphi
-
\frac{C_{\varphi}(\varphi)}{C(\varphi)}a\frac{d(\mathbf{r}/a)}{dt}\dot{\varphi}\
\ \
\end{eqnarray}
where $t$ is the physical time coordinate. If we use the comoving
coordinate $\mathbf{x}$ instead, then this becomes
\begin{eqnarray}
\ddot{\mathbf{x}} + 2\frac{\dot{a}}{a}\dot{\mathbf{x}} &=&
-\frac{1}{a^{3}}\vec{\nabla}_{\mathbf{x}}\Phi  -
\frac{C_{\varphi}(\varphi)}{C(\varphi)}\left(
\frac{1}{a^{2}}\vec{\nabla}_{\mathbf{x}}\varphi +
\dot{\varphi}\dot{\mathbf{x}}\right)
\end{eqnarray}
where we have used Eq.~(\ref{eq:newphi}).

The canonical momentum conjugate to $\mathbf{x}$ is $\mathbf{p}=a^{2}\dot{%
\mathbf{x}}$ so from the equation above we have
\begin{eqnarray}\label{eq:WFdxdtcomov}
\frac{d\mathbf{x}}{dt} &=& \frac{\mathbf{p}}{a^{2}},\\
\label{eq:WFdpdtcomov} \frac{d\mathbf{p}_{\mathrm{CDM}}}{dt} &=&
-\frac{1}{a}\vec{\nabla}_{\mathbf{x}}\Phi -
\frac{C_{\varphi}(\varphi)}{C(\varphi)}\left(\vec{\nabla}_{\mathbf{x}}\varphi+a^2\dot{\varphi}\dot{\mathbf{x}}\right)\nonumber\\
&=& -\frac{1}{a}\vec{\nabla}_{\mathbf{x}}\Phi -
\frac{C_{\varphi}(\varphi)}{C(\varphi)}\left(\vec{\nabla}_{\mathbf{x}}\varphi+\dot{\varphi}\mathbf{p}_{\mathrm{CDM}}\right),\\
\label{eq:WFdpdtcomovb} \frac{d\mathbf{p}_{\mathrm{B}}}{dt} &=&
-\frac{1}{a}\vec{\nabla}_{\mathbf{x}}\Phi,
\end{eqnarray}
where Eq.~(\ref{eq:WFdpdtcomov}) is for CDM particles and Eq.~(\ref%
{eq:WFdpdtcomovb}) is for baryons. Note that according to Eq.~(\ref%
{eq:WFdpdtcomov}) the quantity $a\log [C(\varphi )]$ acts as an effective
potential for the fifth force. This is an important observation and we will
return to it later when we calculate the escape velocity of CDM particles
within a virialized halo.

Eqs.~(\ref{eq:WFphiEOM}, \ref{eq:WFPoisson}, \ref{eq:WFdxdtcomov}, \ref%
{eq:WFdpdtcomov}, \ref{eq:WFdpdtcomovb}) will be used in the code to
evaluate the forces on the dark-matter particles and evolve their positions
and momenta in time.

\subsection{Code Units}

\label{subsect:codeunit}

In our numerical simulation we use a modified version of \texttt{MLAPM} (%
\cite{Knebe:2001}, see \ref{subsect:code} for a brief description of the
code and the essential modifications to it), and we will have to change or
add our Eqs.~(\ref{eq:WFphiEOM}, \ref{eq:WFPoisson}, \ref{eq:WFdxdtcomov}, %
\ref{eq:WFdpdtcomov}, \ref{eq:WFdpdtcomovb}). For this, the first step is to
convert the quantities to the code units of \texttt{MLAPM}. Here, we briefly
summarize the main features.

\texttt{MLAPM} code uses the following internal units (where a subscript $%
_{c}$ stands for "code"):
\begin{eqnarray}
\mathbf{x}_{c} &=& \mathbf{x}/B,\nonumber\\
\mathbf{p}_{c} &=& \mathbf{p}/(H_{0}B)\nonumber\\
t_{c} &=& tH_{0}\nonumber\\
\Phi_{c} &=& \Phi/(H_{0}B)^{2}\nonumber\\
\rho_{c} &=& \rho/\bar{\rho},\nonumber\\
u &=& ac^{2}\sqrt{\kappa}\delta\varphi/\left(H_0B\right)^2,
\end{eqnarray}
where $B$ denotes the comoving size of the simulation box, $H_{0}$
is the present Hubble constant, and $\rho $, with subscript, could
represent the
density of either CDM ($\rho _{c,\mathrm{CDM}}$) or baryons ($\rho _{c,%
\mathrm{B}}$). In the last line the quantity $u$ is the scalar field \emph{%
perturbation }$\delta \varphi $ expressed in terms of code units and is new
to the \texttt{MLAPM} code.

In terms of $u$, as well as the (dimensionless) background value of the
scalar field, $\sqrt{\kappa }\bar{\varphi}$, some relevant quantities are
expressed written in full as
\begin{eqnarray}
V(\varphi) &=&
\frac{\Lambda^{4}}{\left(\sqrt{\kappa}\bar{\varphi}+\frac{B^2H^2_0}{ac^{2}}u\right)^{\alpha}},\nonumber\\
C(\varphi) &=&
\exp\left[\gamma\left(\sqrt{\kappa}\bar{\varphi}+\frac{B^2H^2_0}{ac^{2}}u\right)\right],\nonumber\\
V_{\varphi} &=&
-\alpha\frac{\sqrt{\kappa}\Lambda^{4}}{\left(\sqrt{\kappa}\bar{\varphi}+\frac{B^2H^2_0}{ac^{2}}u\right)^{1+\alpha}},\nonumber\\
C_{\varphi} &=&
\gamma\sqrt{\kappa}\exp\left[\gamma\left(\sqrt{\kappa}\bar{\varphi}+\frac{B^2H^2_0}{ac^{2}}u\right)\right],
\end{eqnarray}
and the background counterparts of these quantities can be obtained simply
by setting $u=0$ (recall that $u$ represents the perturbed part of the
scalar field) in the above equations.

Using all the above newly-defined quantities, we can rewrite Eqs.~(\ref%
{eq:WFphiEOM}, \ref{eq:WFPoisson}, \ref{eq:WFdxdtcomov}, \ref{eq:WFdpdtcomov}%
, \ref{eq:WFdpdtcomovb}) as
\begin{eqnarray}\label{eq:INTdxdtcomov}
\frac{d\mathbf{x}_{c}}{dt_{c}} &=& \frac{\mathbf{p}_{c}}{a^{2}},\\
\label{eq:INTdpdtcomov} \frac{d\mathbf{p}_{c}}{dt_{c}} &=&
-\frac{1}{a}\nabla\Phi_{c} \left[-\frac{1}{a}\frac{C_{\varphi}}{\sqrt{\kappa}C}\left(\nabla u+a\sqrt{\kappa}\dot{\bar{\varphi}}\mathbf{p}_{c}\right)\right],\\
\label{eq:INTPoisson}\nabla^{2}\Phi_{c} &=&
\frac{3}{2}\Omega_{\mathrm{CDM}}\bar{C}
\left(\rho_{c,\mathrm{CDM}}\frac{C}{\bar{C}}-1\right)\nonumber\\
&&+\frac{3}{2}\Omega_{\mathrm{B}}\left(\rho_{c,\mathrm{B}}-1\right)
- \kappa\frac{V-\bar{V}}{H^{2}_{0}}a^{3},
\end{eqnarray}
and
\begin{eqnarray}\label{eq:INTphiEOM}
\nabla^{2}u &=&
\frac{3}{\sqrt{\kappa}}\Omega_{\mathrm{CDM}}\bar{C}_{\varphi}\left(\rho_{c\mathrm{CDM}}\frac{C_{\varphi}}{\bar{C}_{\varphi}}-1\right)
+
\sqrt{\kappa}\frac{V_{\varphi}-\bar{V}_{\varphi}}{H^{2}_{0}}a^{3},\
\ \ \
\end{eqnarray}
where $\Omega _{\mathrm{CDM}}=8\pi G\rho _{\mathrm{CDM}}/3H_{0}^{2}$ and $%
\Omega _{\mathrm{B}}=8\pi G\rho _{\mathrm{B}}/3H_{0}^{2}$ are the dark
matter and baryonic fractional energy densities at the present time. Note
that in Eq.~(\ref{eq:INTdpdtcomov}) the term in the brackets on the
right-hand side only apply to dark matter and not to baryons. Also note that
from here on we shall use $\nabla \equiv \vec{\partial}_{\mathbf{x}%
_{c}},\nabla ^{2}\equiv \vec{\partial}_{\mathbf{x}_{c}}\cdot \vec{\partial}_{%
\mathbf{x}_{c}}$ unless otherwise stated, for simplicity.

We also define
\begin{eqnarray}\label{eq:lambda}
\lambda &\equiv& \frac{\kappa\Lambda^4}{3H_{0}^{2}},
\end{eqnarray}
which will be used frequently below.

Making discrete versions of the above equations for $N$-body simulations is
then straightforward, and we refer the interested readers to Appendix~\ref%
{appen:discret} to the whole treatment, with which we can now proceed to do $%
N$-body simulations.

\section{Simulation Details}

\label{sect:simu}

\subsection{The $N$-Body Code}

\label{subsect:code}

We have modified the publicly available $N$-body code \texttt{MLAPM}
(Multi-Level Adaptive Particle Mesh) for our simulations. This code uses
multilevel grids \cite{Brandt:1977, Press:1992, Briggs:2000} to accelerate
the convergence of the (nonlinear) Gauss-Seidel relaxation method \cite%
{Press:1992} in solving boundary value partial differential equations.
Furthermore, it is also adaptive and refines the grid in regions where the
mass/particle density exceeds a certain predefined threshold. Each
refinement level forms a finer grid (which might contain many parts that are
spatially disconnected) which the particles will be then linked onto and on
which the densities are computed, the scalar field equation and Poisson
equation are solved, the total force on the particles are obtained, and the
particles are drifted and kicked with a smaller time step. Thus \texttt{MLAPM%
} has two kinds of grids: the domain grid which is fixed at the beginning of
a simulation, and refined grids which are generated according to the
particle distribution and which are destroyed after a complete time step.

One benefit of such a setup is that in low-density regions where the
resolution requirement is not high, fewer time steps are needed, while the
bulk of the computing sources can be used in the few high-density regions
where high resolution is needed to ensure precision.

Some technical issues must be controlled. For example, once a refined grid
is created, the particles in that region will be linked to it and densities
on it are calculated, then the coarse-grid values of the gravitational
potential are interpolated to obtain the corresponding values on the finer
grid. When the Gauss-Seidel iteration is performed on refined grids, the
gravitational potential on the boundary nodes is kept constant and only
those values on the interior nodes are updated according to Eq.~(\ref{eq:GS}%
) so as to ensure consistency between coarse and refined grids. This point
is also important in the scalar field simulation because, like the
gravitational potential, the scalar field value is also evaluated on and
communicated between multi-grids (note that different boundary conditions
might lead to different solutions to the scalar field equation of motion).

In our simulation, the domain grid (the finest grid that is not a refined
grid) has $128^{3}$ nodes, and there is a ladder of coarser grids with $%
64^{3}$, $32^{3}$, $16^{3}$, $8^{3}$, $4^{3}$ nodes respectively. These
grids are used for the multi-grid acceleration of convergence: for the
Gauss-Seidel relaxation method, the convergence rate is high for the first
several iterations, but then quickly becomes very slow; this is because the
convergence is only efficient for the high-frequency (short-range) Fourier
modes, while for low-frequency (long-range) modes more iterations do not
help much. To accelerate the solution process, one then switches to the next
coarser grid -- for which the low-frequency modes of the finer grid are
actually high-frequency ones -- and the convergence is speeded up. The
\texttt{MLAPM} Poisson solver adopts the self-adaptive scheme: if
convergence becomes slow on a grid, then go to the next coarser grid where
it is expected to be faster, if convergence is achieved on a grid, then
interpolate the relevant quantities back to the finer grid (provided that
the latter is not on the refinements) and solve the equation there again.
This goes on indefinitely until a converged solution on the domain grid is
obtained, or until one arrives at the coarsest grid (normally with $2^{3}$
nodes) on which the equations can be solved exactly using other techniques,
or by simply iterating many times until convergence is achieved. For the
scalar-field equation of motion, we find that with the self-adaptive scheme
in certain regimes the nonlinear Gauss-Seidel solver tends to fall into
oscillations between the coarser and finer grids; to avoid such situations,
we then use V-cycle \cite{Press:1992} instead.

For the refined grids the method is different. Here, we just iterate Eq.~(%
\ref{eq:GS}) until convergence, without resorting to coarser grids for
acceleration.

As is normal in the Gauss-Seidel relaxation method, convergence is deemed to
be achieved when the numerical solution $u_{n}^{k},$ after $n$ iterations on
grid $k,$ satisfies that the condition that the norm $\Vert \cdot \Vert $
(mean or maximum value on a grid) of the residual
\begin{eqnarray}
e^{k} &=& L^{k}(u^{k}_{n}) - f_{k},
\end{eqnarray}
is smaller than the norm of the truncation error
\begin{eqnarray}
\tau^{k} &=& L^{k-1}(\mathcal{R}u^{k}_{n}) -
\mathcal{R}\left[L^{k}(u^{k}_{n})\right]
\end{eqnarray}
by a certain amount, or, in the V-cycle case, that the reduction
of residual after a full cycle becomes smaller than a predefined
threshold (indeed the former is satisfied whenever the latter is).
Note here that $L^{k}$ is the discretization of the differential
operator Eq.~(\ref{eq:diffop}) on grid $k$ and $L^{k-1}$ a similar
discretization on grid $k-1$, $f_{k}$ is the source term,
$\mathcal{R}$ is the restriction operator to interpolate values
from the grid $k$ to the grid $k-1$. In the modified code we have
used the full-weighting restriction for $\mathcal{R}$.
Correspondingly, there is a prolongation operator $\mathcal{P}$ to
obtain values from grid $k-1$ to grid
$k$, and we use a bilinear interpolation for it. For more details see \cite%
{Knebe:2001}.

\texttt{MLAPM} calculates the gravitational forces on particles by a
centered difference of the potential $\Phi $ and interpolates the forces at
the locations of particles by the so-called triangular-shaped-cloud (\texttt{%
TSC}) scheme to ensure momentum conservation on the grid. The \texttt{TSC}
scheme is also used in the density assignment given the particle
distribution.

Some of our main modifications to the \texttt{MLAPM} code for the coupled
scalar field model are:

\begin{enumerate}
\item We have added a parallel solver for the scalar field, based on Eq.~(%
\ref{eq:u_phi_EOM}). It uses a nonlinear Gauss-Seidel scheme for the
relaxation iteration and the same criterion for convergence as the default
Poisson solver. But is adopts a V-cycle instead of the self-adaptive scheme
in arranging the Gauss-Seidel iterations.

\item The value of $u$ solved thereby is then used to calculate the total
matter density, which completes the calculation of the source term for the
Poisson equation. The latter is then solved using fast Fourier transform on
the domain grids and self-adaptive Gauss-Seidel iteration on refinements.

\item The fifth force is obtained by differentiating the $u$ as the gravity
is computed by differentiating the gravitational potential.

\item The momenta and positions of particles are then updated, or in other
words the particles are \emph{kick}ed and \emph{drift}ed, where in the \emph{%
kick}s we take into account both gravity and the fifth force.
\end{enumerate}

There are a lot of additions and modifications to ensure smooth interface
and the newly added data structures. For the output, as there are multilevel
grids all of which host particles, the composite grid is inhomogeneous and
so we choose to output the positions and momenta of the particles, plus the
gravity, fifth force and scalar field values \emph{at the positions} of
these particles. We can of course easily read these data into the code,
calculate the corresponding quantities on each grid and output them if
needed. We also output the potential and scalar field values on the $128^{3}$
domain grid.

\subsection{Physical and Simulation Parameters}

\label{subsect:param}

The physical parameters we use in the simulations are as follows: the
present-day dark-energy fractional energy density $\Omega _{\mathrm{DE}%
}=0.743$ and $\Omega _{m}=\Omega _{\mathrm{CDM}}+\Omega _{\mathrm{B}}=0.257$%
, $H_{0}=71.9$~$km/s/Mpc$, $n_{s}=0.963$, $\sigma _{8}=0.761$. Our
simulation box has a size of $64h^{-1}$~Mpc, where $h=H_{0}/(100~\mathrm{%
km/s/Mpc})$. We simulate 4 models, with parameters $\alpha =0.1$ and $\gamma
=-0.05,-0.10,-0.15,-0.20$ respectively (such parameters are chosen so that
the deviation from $\Lambda $CDM will not become too large to be realistic).
In all those simulations, the mass resolution is $1.114\times
10^{9}h^{-1}~M_{\bigodot }$; the particle (both dark matter and baryons)
number is $256^{3}$ (see Sect.~\ref{subsect:baryons_inc} for a discussion);
the domain grid is a $128\times 128\times 128$ cubic and the finest refined
grids have 16384 cells on each side, corresponding to a force resolution of $%
\ $about $12h^{-1}~$kpc.

We also make a run for the $\Lambda$CDM model using the same physical
parameters and different initial condition (see Sect.~\ref{subsect:ic}).

\begin{figure*}[tbp]
\centering \includegraphics[scale=1.05] {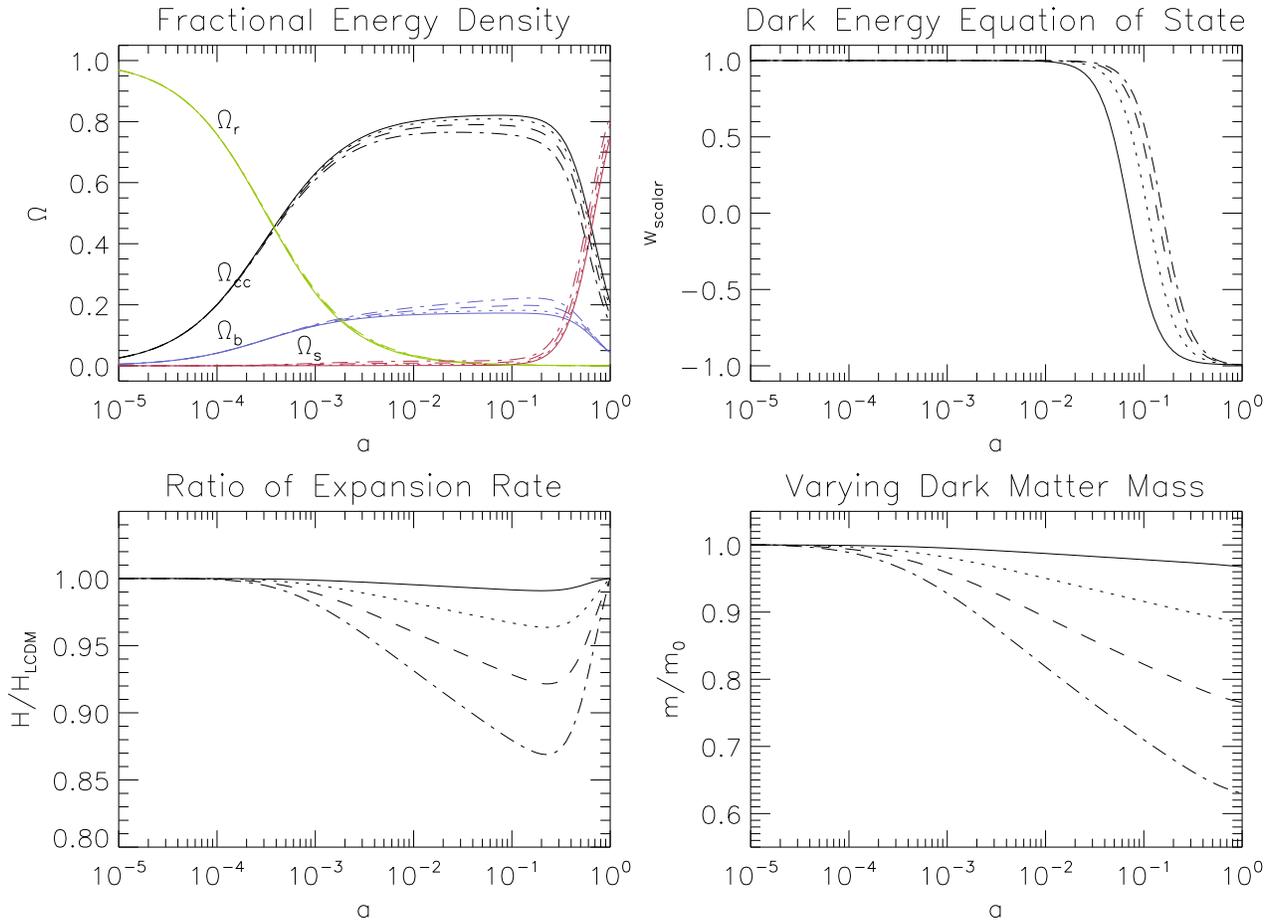}
\caption{(Color Online) Figures to illustrate the background evolution of
our coupled scalar field models. \emph{Upper Left Panel}: The fractional
energy densities for radiation (green), \emph{coupled} dark matter (black),
baryons (blue) and the scalar field (red); note that $\Omega_{\mathrm{cc}}=8%
\protect\pi GC(\protect\varphi)\protect\rho_{\mathrm{CDM}}/3H^2$. \emph{%
Upper Right Panel}: The equation of state of the scalar field, $w\equiv p_{%
\protect\varphi}/\protect\rho_{\protect\varphi}$ in which $p_{\protect\varphi%
}=\frac{1}{2}\dot{\protect\varphi}^2-V(\protect\varphi)$ and $\protect\rho_{%
\protect\varphi}=\frac{1}{2}\dot{\protect\varphi}^2+V(\protect\varphi)$.
\emph{Lower Left Panel}: The ratio between the Hubble expansion rate in the
coupled scalar field model and that in the $\Lambda$CDM paradigm, other
physical parameters such as $\protect\rho_{\mathrm{B}}, \protect\rho_{%
\mathrm{RAD}}, \protect\rho_{\mathrm{CDM}}$ being held the same, as a
function of the scale factor $a$. \emph{Lower Right Panel}: The "varying
mass" of the dark matter particles as a function of $a$ -- here $m_0$ is the
\emph{constant bare} mass of the particles and $m=e^{\protect\gamma\protect%
\varphi}m_0$. In all figures we have chosen $\protect\alpha=0.1$ and the
solid, dotted, dashed and dot-dashed curves represent the models with $%
\protect\gamma=-0.05, -0.10, -0.15$ and $-0.20$ respectively.}
\label{fig:background}
\end{figure*}

\begin{figure*}[tbp]
\centering \includegraphics[scale=1.05] {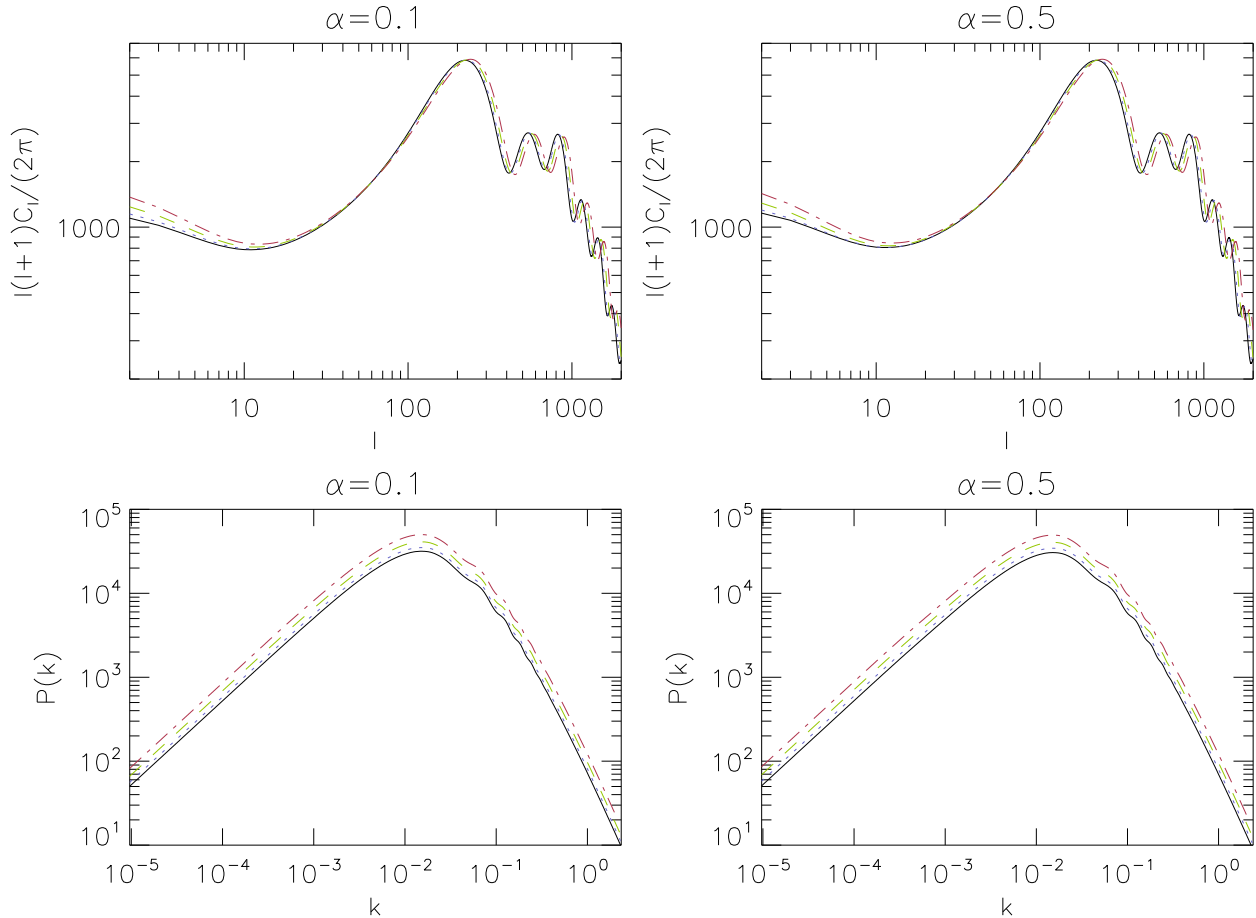}
\caption{(Color Online) The linear power spectra of the coupled scalar field
model. \emph{Upper Left Panel}: The Cosmic Microwave Background (CMB) power
spectra for the four models with $\protect\alpha=0.1$. \emph{Upper Right
Panel}: The same but for the models with $\protect\alpha=0.5$. \emph{Lower
Left Panel}: The matter power spectra at present day (redshift $z=0$) for
the four models with $\protect\alpha=0.1$. \emph{Lower Right Panel}: The
same but for the models with $\protect\alpha=0.5$. In all figures the solid
(black), dotted (blue), dashed (green) and dot-dashed (red) curves represent
the models with $\protect\gamma=-0.05, -0.10, -0.15$ and $-0.20$
respectively.}
\label{fig:linpert}
\end{figure*}

\begin{figure*}[tbp]
\centering \includegraphics[scale=1.05] {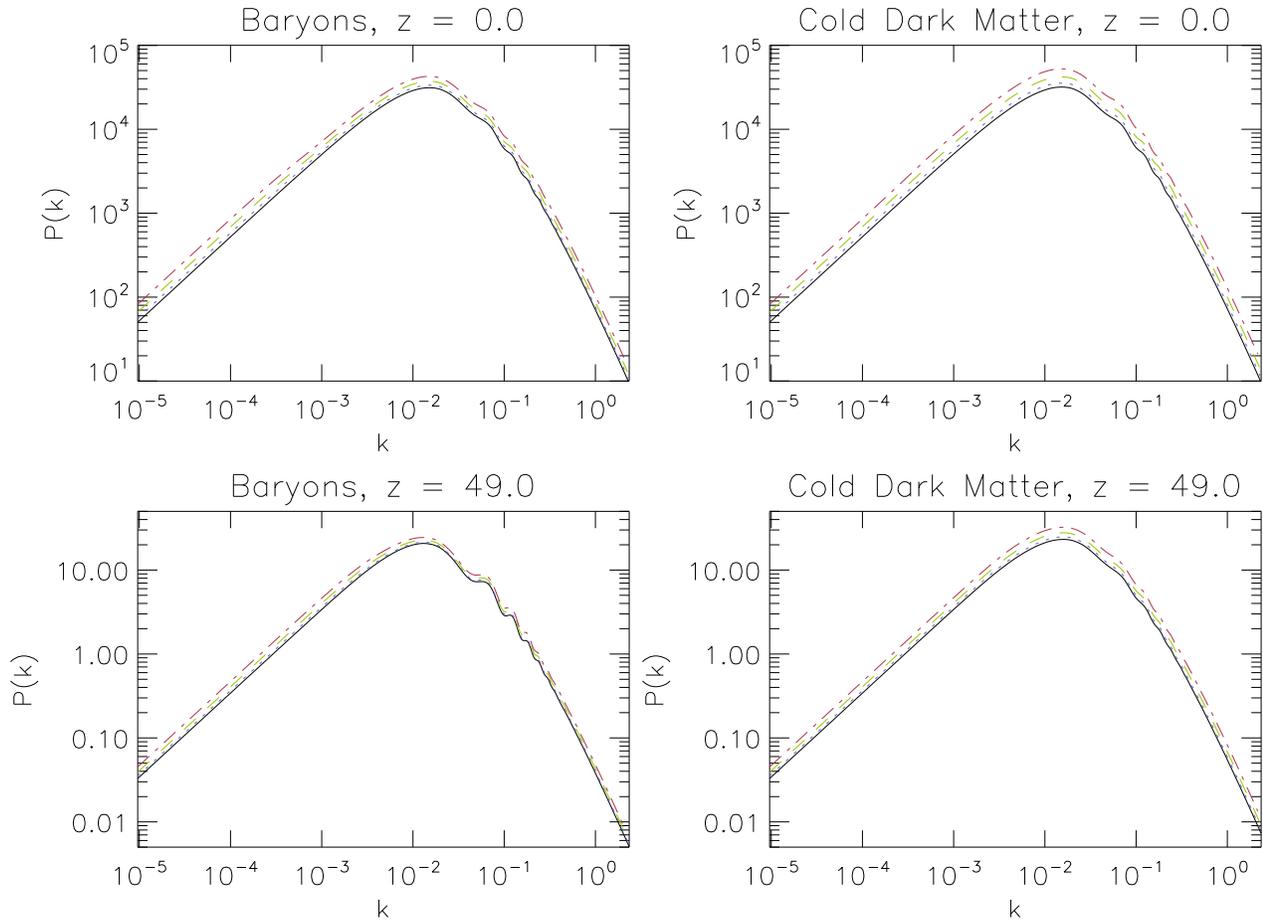}
\caption{(Color Online) \emph{Upper Panels}: The matter power spectra for
baryons only (\emph{Left}) and for dark matter only (\emph{Right}), at
current time ($z=0$). \emph{Lower Panels}: The same but at an earlier time $%
z=49$ where the initial condition for the $N$-body simulations are computed.
In all figures $\protect\alpha=0.1$ and the solid (black), dotted (blue),
dashed (green) and dot-dashed (red) curves represent the models with $%
\protect\gamma=-0.05, -0.10, -0.15$ and $-0.20$ respectively.}
\label{fig:linpert2}
\end{figure*}

\begin{figure*}[tbp]
\centering \includegraphics[scale=1.05] {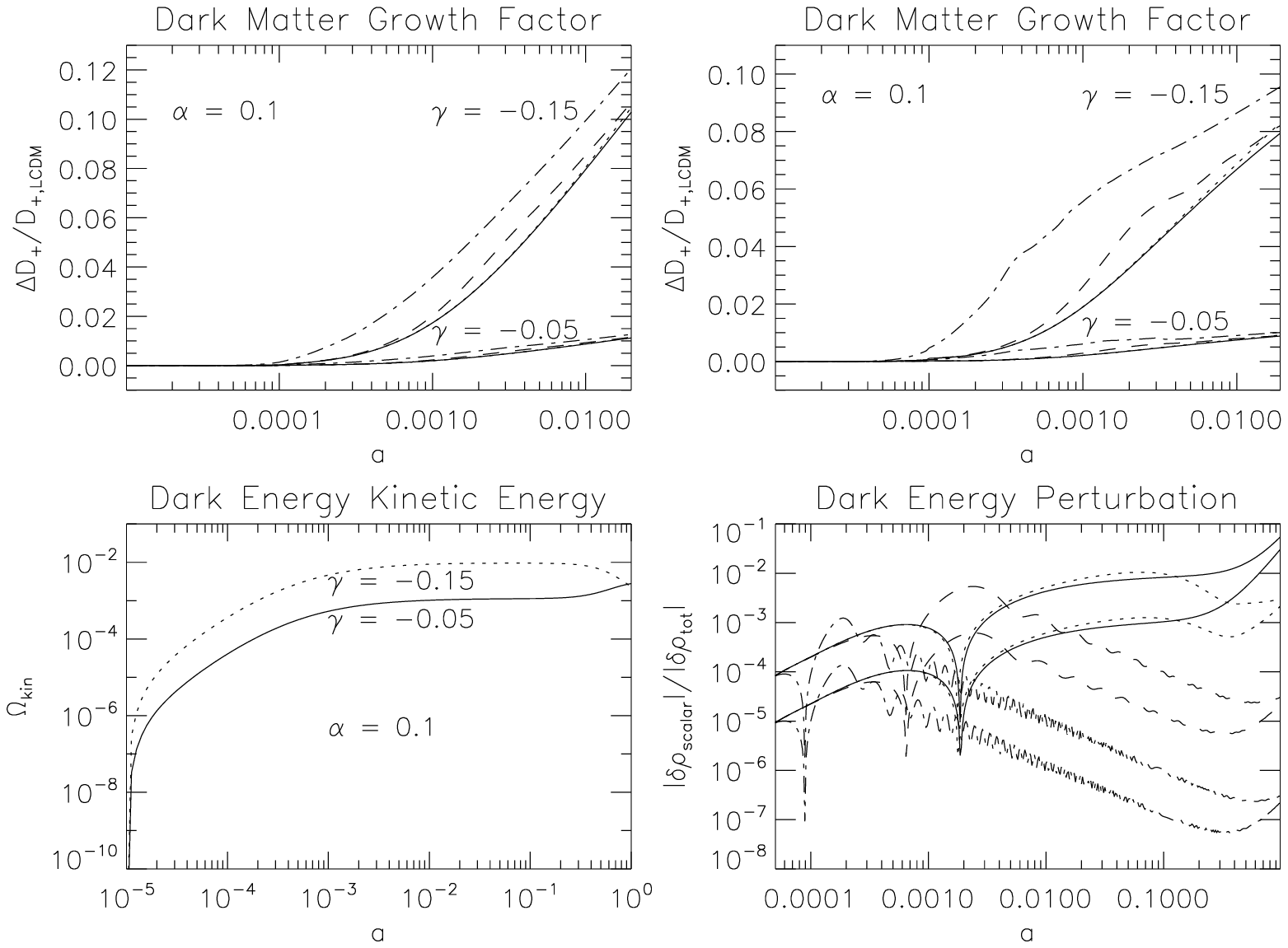}
\caption{\emph{Upper Left Panel}: The fractional change of the growth factor
$D_{+}$ of the dark matter density perturbation in the coupled scalar field
model as compared with the $\Lambda$CDM prediction; for clearness shown are
only the results for the two models with $\protect\alpha=0.1,\protect\gamma%
=-0.05$ and $\protect\alpha=0.1,\protect\gamma=-0.15$, as indicated above
the curves, and for each model the solid, dotted, dashed and dot-dashed
curves represent respectively the result for $k=0.0001, 0.001, 0.01$ and $%
0.1~\mathrm{Mpc}^{-1}$. \emph{Upper Right Panel}: The same as above, but for
the fractional change of $\dot{D}_{+}$ as a function of the scalar factor $a$%
. \emph{Lower Left Panel}: The time evolution of the fractional energy
density of the \emph{kinetic energy} of the scalar field, $\Omega_{\mathrm{%
kin}}=4\protect\pi G\dot{\protect\varphi}^2/3H^2$ for the two models with $%
\protect\alpha=0.1,\protect\gamma=0.05$ (solid curve) and $\protect\alpha%
=0.1,\protect\gamma=-0.15$ (dotted curve) respectively. \emph{Lower Right
Panel}: The evolution of the density contrast of the scalar field, of which
the energy density is $\protect\rho_{\protect\varphi}=\frac{1}{2}\dot{%
\protect\varphi}^2+V(\protect\varphi)$; the solid, dotted, dashed and
dot-dashed curves are for four different length scales $k=0.0001, 0.001,
0.01 $ and $0.1~\mathrm{Mpc}^{-1}$ respectively, and for each style of curve
the upper (lower) one is for the model with $\protect\alpha=0.1,\protect%
\gamma=-0.15$ ($\protect\alpha=0.1,\protect\gamma=-0.05$).}
\label{fig:dplus}
\end{figure*}

\subsection{Distinguishing Baryons from Dark Matter}

\label{subsect:baryons_inc}

In our simulations only dark matter particles are coupled to the scalar
field. The baryons do not contribute to the scalar field equation of motion,
nor are they influenced by the scalar-field fifth force. Therefore, it is
important to make sure that they are distinguished and treated appropriately.

In the modified code we distinguish baryons and dark matter particles by
tagging them differently. We consider the situation where 17.12\% of all our
matter particles are baryonic ($\Omega _{\mathrm{B}}=0.044$) and 82.88\% are
dark matter \footnotemark[1]  . During the generation of the initial
condition (initial distribution and displacements of particles), we loop
over all particles and for each particle we generate a random number from a
uniform distribution in $[0,1]$. If this random number is less than 0.1712
then we tag the particle as baryon, otherwise we tag it as dark matter. Once
these tags have been set up they are never changed again, and the code then
determines whether\ or not a particle contributes to the scalar field
evolution and feels the fifth force according to its tag.

\footnotetext[1]{%
It is time to stress that our $\Omega_{\mathrm{CDM}}=8\pi G\rho_{\mathrm{CDM}%
}/3H_0^2$ is the fractional energy density of the bare dark matter
particles, which is \emph{not} weighed by the coupling function $C(\varphi)$%
. As we mentioned in Appendix~\ref{appen:bkgd} and \cite{Li:2009sy}, $\rho_{%
\mathrm{CDM}}\propto a^{-3}$ as in $\Lambda$CDM but the behaviour of $%
C(\varphi)\rho_{\mathrm{CDM}}$ (which is the actual quantity appearing in
the Poisson equation) can be rather complicated. In all models we are
simulating, including the $\Lambda$CDM one, we have the same $\rho_{\mathrm{%
CDM}}$, \emph{rather than} $C(\varphi)\rho_{\mathrm{CDM}}$, at present-day.}

\begin{figure*}[tbp]
\centering \includegraphics[scale=0.7] {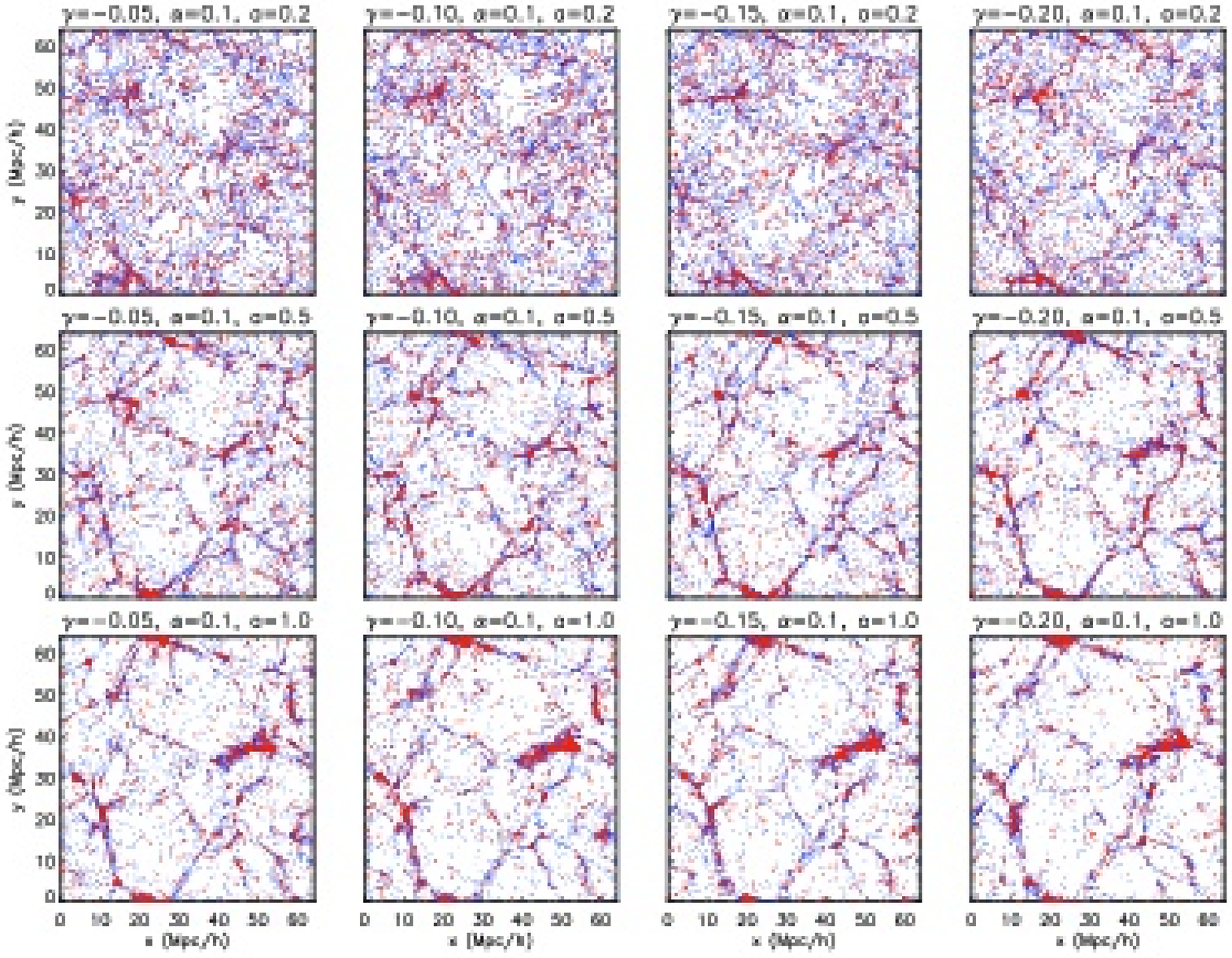}
\caption{(Color Online) Snapshots of the particle distribution in
our four coupled scalar field models as indicated by the subtitles
of the panels. $a$ is the scale factor and $a=1$ is the present
time. For clearness we only
pick out a slice of the simulation box with $30~h^{-1}\mathrm{Mpc}%
<z<30.3~h^{-1}\mathrm{Mpc}$ and $0~h^{-1}\mathrm{Mpc}<x,y<64~h^{-1}\mathrm{%
Mpc}$. The blue dots represent dark matter particles and red dots baryons.}
\label{fig:particle}
\end{figure*}

\begin{figure*}[tbp]
\centering \includegraphics[scale=0.7] {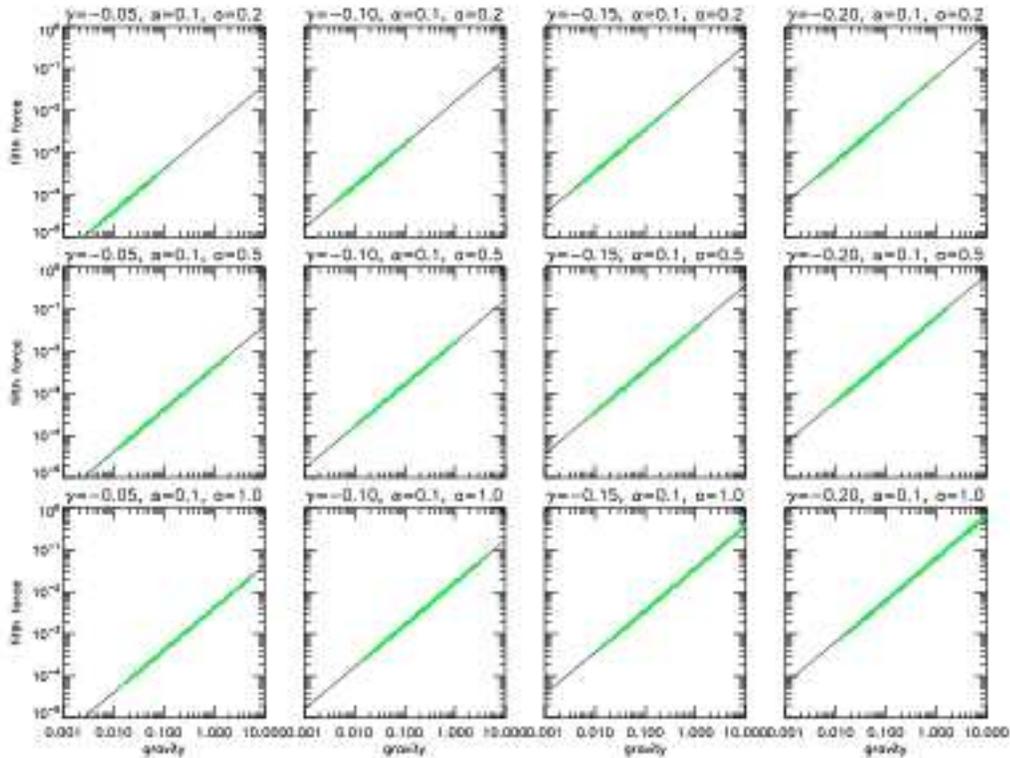} \caption{(Color
Online) The relation between the magnitudes of the fifth force and
that of gravity for the four coupled scalar field models at three
different output times $a=0.2, 0.5, 1.0$. The black solid line in
each panel
represents the analytical approximation $f=2\protect\gamma^2\cdot\frac{%
\Omega_{\mathrm{CDM}}}{\Omega_{m}}g$ (see text) and the $\sim150,000$ green
dots the results from the simulations. The particles are the same as those
in Fig.~\protect\ref{fig:particle}.}
\label{fig:force}
\end{figure*}

\begin{figure*}[tbp]
\centering \includegraphics[scale=0.7] {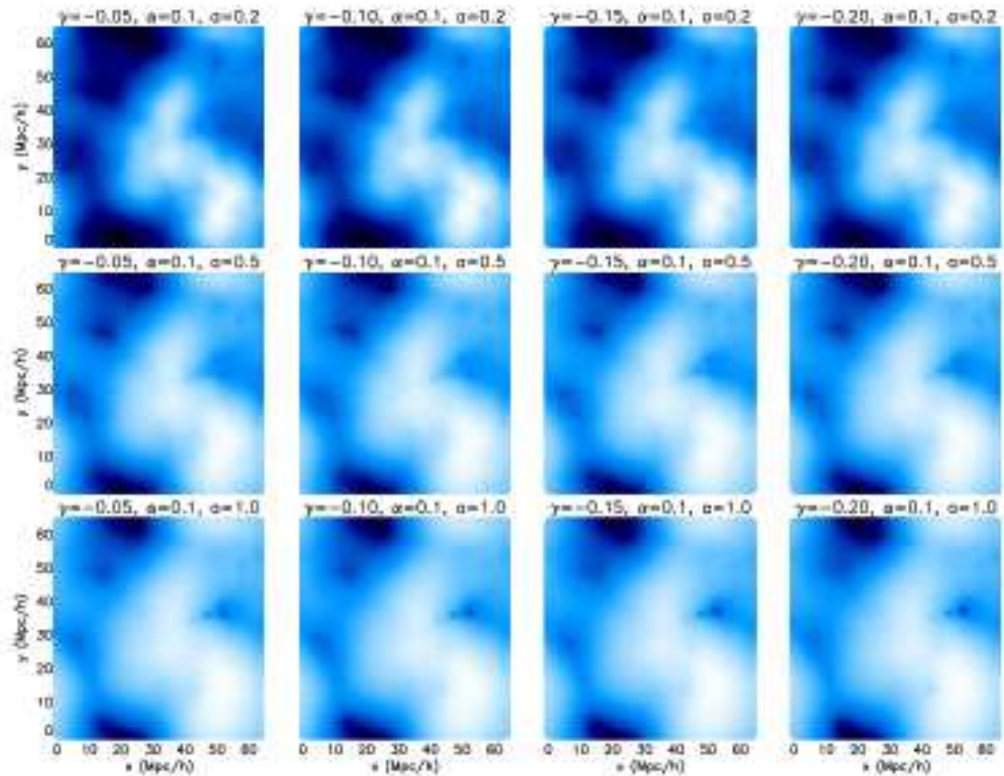}
\caption{(Color Online) The gravitational potential on the $z$ plane where $%
z=32h^{-1}$~Mpc. Dark regions are where the potential is deeper while light
regions are where it is shallower. The four columns are for the four models
we consider and the three rows are for three output times, respectively $%
a=0.2, 0.5$ and $1$ where $a$ is the cosmic scale factor.}
\label{fig:potential}
\end{figure*}

\begin{figure*}[tbp]
\centering \includegraphics[scale=0.7] {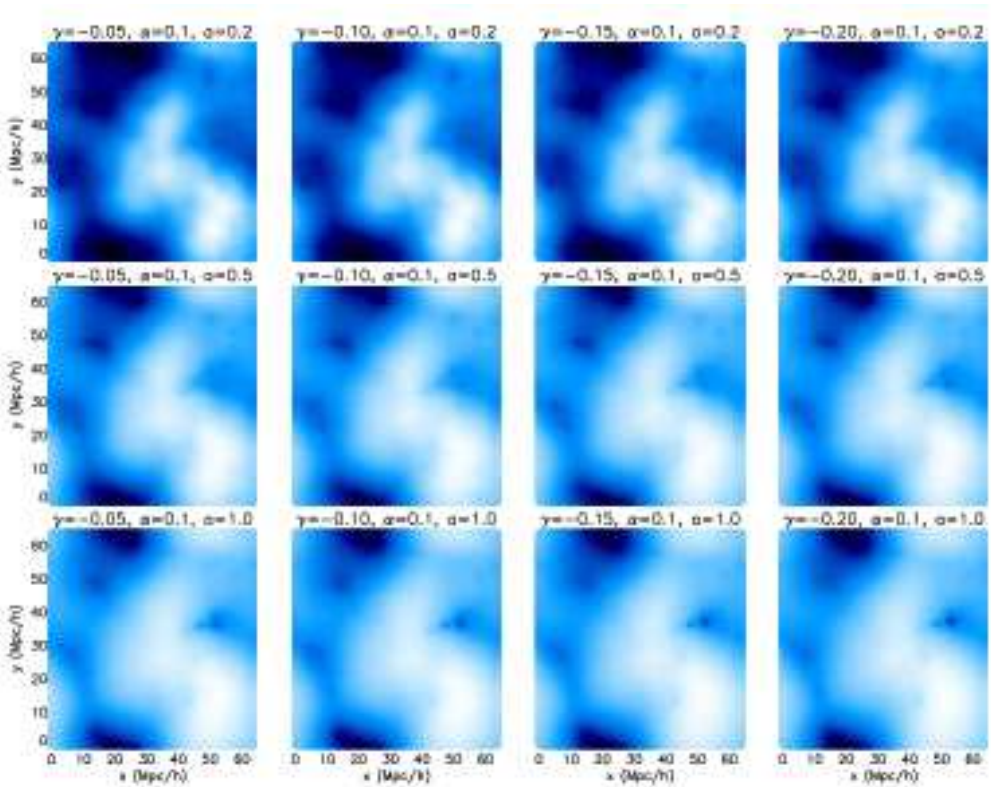}
\caption{(Color Online) The same as
Fig.~\protect\ref{fig:potential} but for $-\protect\varphi$ where
$\protect\varphi$ is the value of the scalar field.
The negative sign is added to make the plot look similar to Fig.~\protect\ref%
{fig:potential}, and otherwise they will be compensating for each other.}
\label{fig:scalar}
\end{figure*}

\begin{figure*}[tbp]
\centering \includegraphics[scale=1.9] {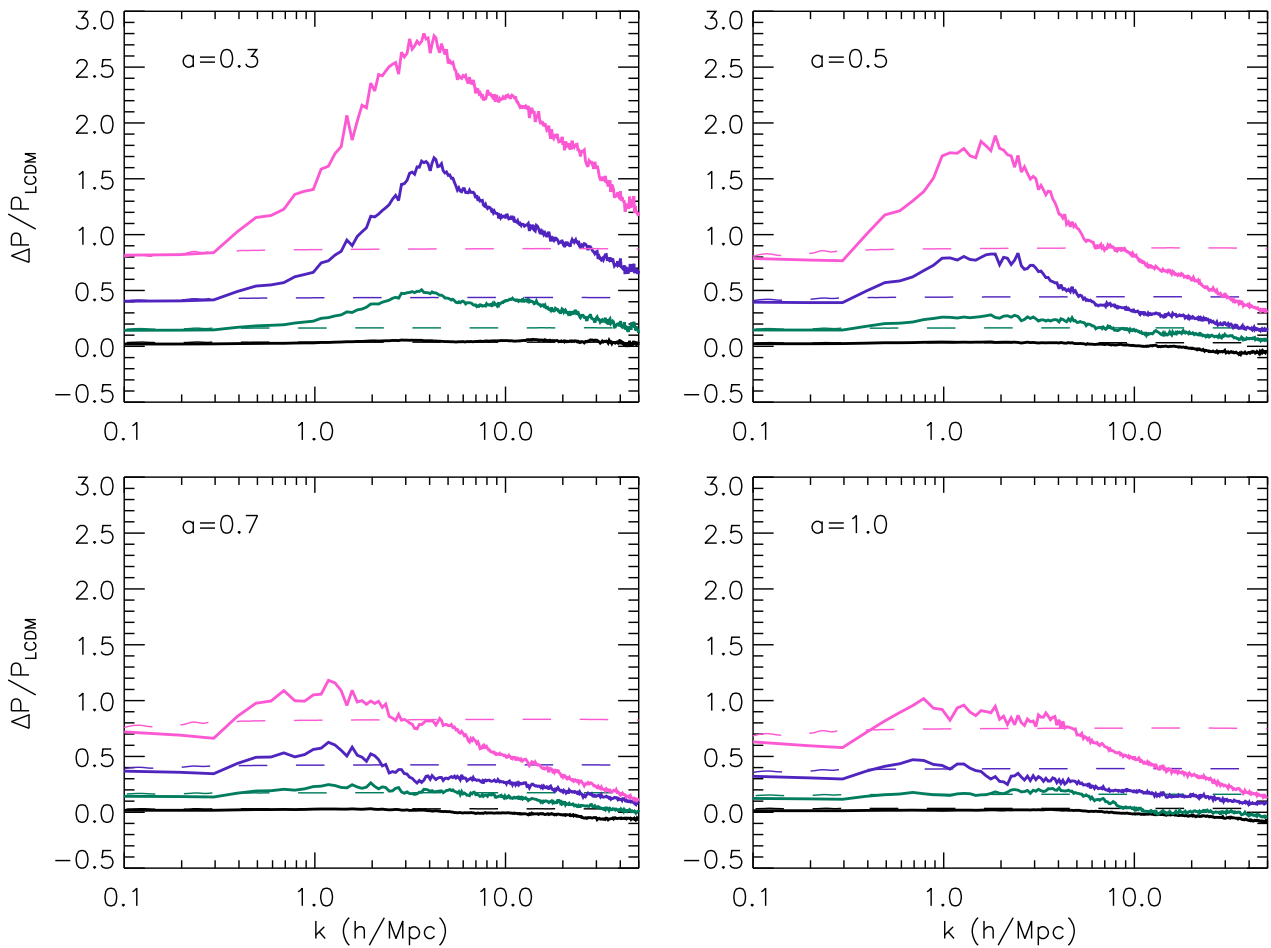}
\caption{(Color Online) The linear and nonlinear matter power spectra $P(k)$
as compared with those of the $\Lambda$CDM paradigm. Shown are the
fractional change of $P(k)$ for the four models with $\protect\alpha=0.1$
and $\protect\gamma=-0.05$ (black curves), $-0.10$ (green), $-0.15$ (blue), $%
-0.20$ (purple) and at four different output times $a=0.3$ (\emph{Upper Left
Panel}), $0.5$ (\emph{Upper Right}), $0.7$ (\emph{Lower Left}), $1.0$ (\emph{%
Lower Right}). Solid curves are from $N$-body simulations and dashed curves
from linear perturbation calculation.}
\label{fig:power}
\end{figure*}

\begin{figure*}[tbp]
\centering \includegraphics[scale=1.9] {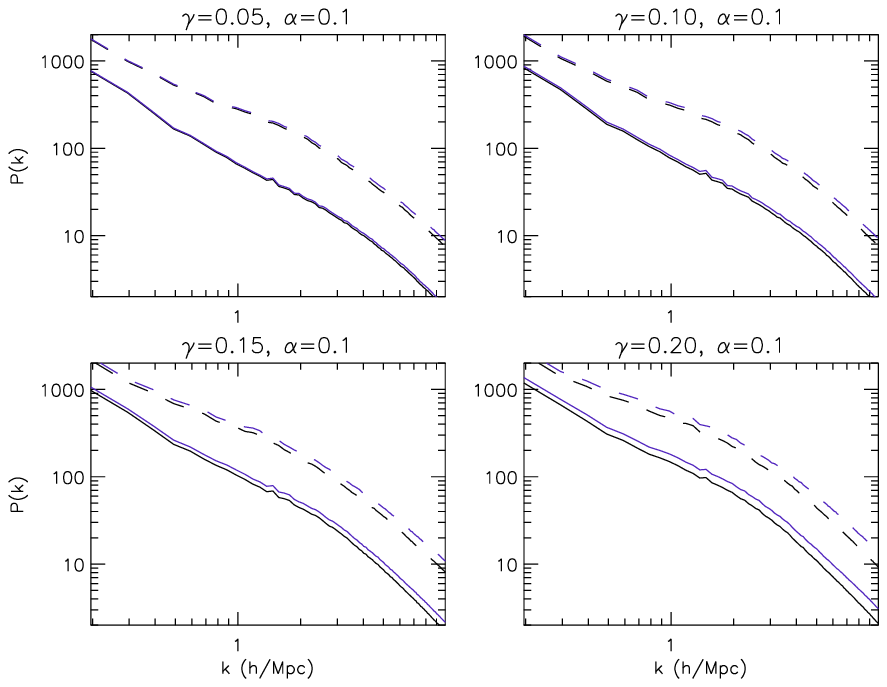}
\caption{(Color Online) The bias between the nonlinear matter power spectra $%
P(k)$ of dark matter and baryons for the four models we are simulating, as
indicated in the subtitles of the panels. The solid curves are for the
output time $a=0.5$ and dashed curves for $a=1.0$. For each style of curve
the lower (black) one is for baryons and the upper (purple) one is for dark
matter.}
\label{fig:powerbc}
\end{figure*}

Often in $N$-body simulations, the number of particles is chosen to be the
same as the number of cells in the domain grid, but in our simulations we
have set the former ($256^{3}$) to be eight times bigger than the latter ($%
128^{3}$). This choice obviously increases the mass resolution, but what is
more important for us is that it produces a smoother and more reasonable
dark matter density on the grid. In order to see this, remember that only $%
82.88\%$ of all particles are dark matter, which means that, if we use $%
128^{3}$ rather than $256^{3}$ particles, about $1/5$ of all the grid cells
do not contain dark matter particles at the initial time. The resulting dark
matter density therefore shows some artificial discreteness, which not only
does not reflect reality but might also cause the solver for the scalar
field equation to diverge. By having more particles, we can make the dark
matter densities smoother and resolve the problem of over-discreteness.

\subsection{Background and Linear Perturbation Evolution}

\label{subsect:bkgd}

In general, a coupling between the scalar field and the dark matter
particles not only affects the force law and clustering properties of those
particles [as described by Eqs.~(\ref{eq:WFphiEOM}, \ref{eq:WFPoisson}, \ref%
{eq:WFdxdtcomov}, \ref{eq:WFdpdtcomov}, \ref{eq:WFdpdtcomovb})], but it also
influences the background and linear perturbation evolution of the Universe.
The modification in the background cosmic expansion rate directly changes
the rate of the matter clustering, while the modification in the linear
perturbation growth might lead to a different initial condition for the $N$%
-body simulation from the $\Lambda $CDM result. Consequently, we must take
appropriate care of these issues in the $N$-body simulations in order to
obtain reliable and complete numerical results.

The background expansion rate of the universe is completely governed by (the
zeroth order parts of) the scalar field equation of motion
\begin{eqnarray}\label{eq:bkgd_scalar}
\ddot{\varphi} + 3H\dot{\varphi} + V_{\varphi} +
C_{\varphi}\rho_{\mathrm{CDM}} &=& 0,
\end{eqnarray}
the Friedman equation
\begin{eqnarray}\label{eq:bkgd_friedman}
3H^2 &=& \kappa\left[\rho_{\mathrm{B}}+\rho_{\mathrm{RAD}}
+C(\varphi)\rho_{\mathrm{CDM}}+\frac{1}{2}\dot{\varphi}^{2}
+V(\varphi)\right]\ \ \
\end{eqnarray}
where $H=\dot{a}/a$ is the Hubble expansion rate and $\rho _{\mathrm{RAD}}$
includes contributions from photons and massless neutrinos (\emph{i.e.},
radiation), and the Raychaudhuri equation
\begin{eqnarray}\label{eq:bkgd_raychaudhuri}
&& 3\left(\dot{H}+H^2\right)\nonumber\\ &=&
-\frac{\kappa}{2}\left[\rho_{\mathrm{B}}+2\rho_{\mathrm{RAD}}
+C(\varphi)\rho_{\mathrm{CDM}}+2\dot{\varphi}^{2}
-2V(\varphi)\right].\ \ \
\end{eqnarray}

The introduction of Eq.~(\ref{eq:bkgd_scalar}) introduces a new degree of
freedom $\varphi $ to the system, for which the initial condition and
relevant parameter (the $\Lambda $ in $V(\varphi )$) should be chosen
appropriately to guarantee consistency. More explicitly, we have to adjust
the values of $\Lambda $ and $\varphi _{\mathrm{ini}},\dot{\varphi}_{\mathrm{%
ini}}$ so that today $H=H_{0}$ where $H_{0}$ is the measured Hubble constant
which we set to be $71.9~\mathrm{km/s/Mpc}$ in this work. This is a
nontrivial requirement and is achieved through a trial-and-error process. We
shall leave all the details of the numerical algorithm to Appendix~\ref%
{appen:bkgd}, as they are not our major concern here.

In Fig.~\ref{fig:background} we have shown some representative results for
the influences of the scalar field coupling on the background cosmology.
Clearly, for fixed $\alpha $, increasing $|\gamma |$ means increasing the
coupling strength, leading to more dramatic evolution of the scalar field,
and this is why in the lower right panel we can see that as $|\gamma |$
increases, so the quantity $e^{\gamma \varphi }$ falls increasingly below $1$%
. Meanwhile, as the scalar field rolls faster for larger $|\gamma |$, the
equation of state parameter $w$ approaches $-1$ (potential-energy-dominated
regime) later in time (the upper right panel). Also, because $e^{\gamma
\varphi }$ deviates more from $1$, the contribution of the coupled dark
matter ($\rho _{\mathrm{cc}}=\rho _{\mathrm{CDM}}e^{\gamma \varphi }$ in
which $\rho _{\mathrm{CDM}}\propto a^{-3}$, see Appendix~\ref{appen:bkgd})
to the Friedmann equation [Eq.~(\ref{eq:bkgd_friedman})] is smaller and the
contributions from other matter species are greater (upper left panel).
Because dark matter is the principal ingredient driving the expansion of the
Universe in the matter-dominated era, this in turn implies that the
expansion rate will differ from the $\Lambda $CDM prediction (lower left
panel). Note that the change of expansion rate at early times necessarily
cause the matter power spectrum at $z\sim 50$ to differ from the $\Lambda $%
CDM result, a point we will return to shortly.

In Fig.~\ref{fig:linpert} we have displayed the effects of the scalar field
coupling on the linear cosmic microwave background (CMB) and matter power
spectra. Because of the decrease in the expansion rate, the angular diameter
distance increases so that the peaks of the CMB spectrum are shifted
rightwards, and on large scales the perturbation in the scalar field
(cf.~Fig.~\ref{fig:dplus} below) modifies the integrated Sachs-Wolfe effect
to increase the power at low $\ell $. The matter power spectrum, on the
other hand, is increased on all scales: on large scales this is purely
because the universe is now expanding more slowly, allowing matter to
cluster more. On small scales there is another effect -- the boost due to
the scalar field fifth force, which is almost always proportional to gravity
in magnitude and parallel in direction (see below). It is clear from the
figure that the results are rather insensitive to the parameter $\alpha $,
and so in what follows we shall only consider the case of $\alpha =0.1$.

Since we need to know the matter power spectrum at early time $z\sim 50$ in
order to general initial conditions for the $N$-body simulations, we also
plot it in Fig.~\ref{fig:linpert2}. We further separate the baryons from
dark matter, the latter being our major concern: on small scales it is clear
that the dark matter power spectrum is greater than that of baryons, due to
the extra fifth force it experiences (cf.~upper panels). Most interestingly,
we see that even at $z=49$ the dark matter power spectra for different
models can be rather different, again due to the modified expansion rate and
the fifth force. This means that in our $N$-body simulations we should \emph{%
not} use the same initial condition (as in \cite{Li:2009sy, Li:2010}, where
the chameleon effect is so strong that at early times the matter power
spectrum is indistinguishable from that of $\Lambda $CDM). Instead, we
should generate initial conditions separately for different models (in our
case different values for $\gamma $) -- this is the topic of Sect.~\ref%
{subsect:ic} and Appendix~\ref{appen:discret}.

\subsection{Initial Conditions for the $N$-Body Code}

\label{subsect:ic}

As we described in detail above, the fact that the
scalar-field-dark-matter coupling begins to take effect at rather
high redshift indicates that the initial conditions for the N-body
simulations of our coupled scalar field models (at redshift
$z_{i}\sim 50$ here) will also be different from those in $\Lambda
$CDM. To account for this modification, the most straightforward
approach is to generate the linear matter power spectra for our
coupled scalar field models at the starting redshift $z_{i}\sim
50$, and utilize them to produce the Gaussian random density
fluctuation field and displace the particles. In \texttt{GRAFIC2}
the matter power spectrum is generated at time $z=0$, normalized
to a pre-selected value and then evolved back to $z_{i}$ and used
to displace particles. We shall not follow this in our work;
instead we adopt the same initial condition in
\texttt{CAMB} at redshift $z\sim 10^{6}$ for all models including the $%
\Lambda $CDM one and evaluate the power spectrum at $z_{i}$ -- in
this way the linear $\sigma_{8}$ today will be different for
different models (in contrast $\sigma_{8}$ is the same for all
models in some other works). This is because we are interested in
how the scalar field affects the evolution of structure as
compared to $\Lambda $CDM, given the same initial condition at
very early times.

In principle, the matter power spectrum at $z_{i}$ that is used in \texttt{%
GRAFIC2} needs to be generated for each model, for example by linear
perturbation code that is default in \texttt{GRAFIC2}; however, here we
adopt a more economical method. To see this, in the upper panels of Fig.~\ref%
{fig:dplus} we have displayed the time evolution of $\left[
D_{+}-D_{+,\Lambda \mathrm{CDM}}\right] /D_{+,\Lambda \mathrm{CDM}}$ as well
as $\left[ \dot{D}_{+}-\dot{D}_{+,\Lambda \mathrm{CDM}}\right] /\dot{D}%
_{+,\Lambda \mathrm{CDM}}$, with $D_{+}$ being the linear growth factor we
have discussed in Appendix~\ref{appen:Zeldovich}. For simplicity, we only
show these for the two models with $\alpha =0.1$ and $\gamma =-0.05,-0.15$
as indicated in the figure, and for each model the solid, dotted, dashed and
dash-dotted curves represent respectively the results for $k=0.0001$, $0.001$%
, $0.01$ and $0.1\mathrm{Mpc}^{-1}$. There are several important features in
these plots. First, at early times ($z\gtrsim 10^{4}$) we see that the
difference is negligible as the scalar field has yet to take effect. Second,
on large scales (small $k$) the results tend to converge to a curve which
describes how the change of background expansion rate modifies the density
growth (the fifth force has no effect here because the scale is out of its
range). Third, on smaller scales (large $k$) the fifth force begins to act
and further enhances the growth of dark matter density perturbation. These
plots show that $\tilde{D}_{+}$, the dark matter linear growth factor,
depends on both $k$ and time as we mentioned in Appendix~\ref%
{appen:Zeldovich}, and this must be taken into account.

As detailed in Appendix~\ref{appen:Zeldovich}, the displacement and peculiar
velocity of particles are given by $D_{+}\mathbf{d}$ and $\dot{D}_{+}\mathbf{%
d}$ respectively, where $\mathbf{d}$ is a vector which is the same for
different models (remember again we start with the same initial condition in
\texttt{CAMB} at $z\sim 10^{6}$). Consequently, the differences in the
displacements and peculiar velocities in our models are equivalent to the
differences in $D_{+}\left( z_{i}\right) $ and $\dot{D}_{+}\left(
z_{i}\right) $. Instead of evaluating $D_{+}\left( z_{i}\right) $ and $\dot{D%
}_{+}\left( z_{i}\right) $ with (a modified version of) \texttt{GRAFIC2} for
each model, we use the $\Lambda $CDM results for $D_{+}\left( z_{i}\right) $
and $\dot{D}_{+}\left( z_{i}\right) $ computed by \texttt{GRAFIC2}, and also
calculate $\mu _{1}\equiv D_{+}/D_{+,\Lambda \mathrm{CDM}}$ and $\mu
_{2}\equiv \dot{D}_{+}/\dot{D}_{+,\Lambda \mathrm{CDM}}$ at $z_{i}$ using
\texttt{CAMB}. Then the particle displacement and peculiar velocity in the
coupled scalar field models are simply given respectively by $\mu _{1,2}$
times those generated by \texttt{GRAFIC2}.

Furthermore, we shall take into account the differences between the coupling
and non-coupling matter species as following. The baryons do not feel the
fifth force (remember that 'baryons' here simply means non-coupling dark
matter particles, and this is why we do not use the baryon matter power
spectrum generated by \texttt{CAMB} to displace these particles). The
difference between the baryonic linear growth factor $D_{+}$ from $%
D_{+,\Lambda \mathrm{CDM}}$ is mainly due to the modified expansion rate,
and so we use the $D_{+}$ calculated for very small $k$ to evaluate $\mu
_{1,2}$ for baryons. Dark matter particles, on the other hand, do feel the
fifth force which, on small scales, has a magnitude of $2\gamma ^{2}$ times
that of gravity, and so we use the $\tilde{D}_{+}$ calculated for large $k$
to evaluate $\tilde{\mu}_{1,2}$ for dark matter. In this way, the bias
between the two matter species is approximately reflected in our initial
conditions. The scale dependence of $D_{+}$ does not appear to have been
considered in generating initial conditions in the earlier literature.

When generating the initial displacement and peculiar velocities of
particles, we run a random number generator for a variable with a uniform
distribution $U[0,1]$. As mentioned above, if the number generated is less
than $\Omega _{\mathrm{B}}/\left( \Omega _{\mathrm{B}}+\Omega _{\mathrm{CDM}%
}\right) $, we tag the particle as a baryon and use $\mu _{1,2}$ to displace
it; otherwise we tag the particle as dark matter and use $\tilde{\mu}_{1,2}$
to compute its displacement and velocity. Once this has been done, no
particle will ever be re-tagged so that the consistency is not spoiled.
Those particles tagged as dark matter will contribute to the scalar field
evolution and be influenced by the scalar field fifth force, while those
which are tagged as baryons will not.

\section{Simulation Results}

\label{sect:results}

In this section we present the main results from the $N$-body simulations we
have performed. We start by listing some preliminary results to give a rough
idea about the effects of the couplings we are studying.

\subsection{Preliminary Results}

\label{subsect:snapshots}

In Fig.~\ref{fig:particle} we show snapshots of the particle distribution at
different output times. We can see clearly the trend of matter clustering.
The blue and red dots are the dark matter and baryons particles respectively.

Since one of the most important influences of the scalar field coupling is
to exert a fifth force on the dark matter particles, we are interested in
the magnitude of the fifth force as compared with that of gravity. This is
given in Fig.~\ref{fig:force}, where we have again displayed the results for
the simulated models at different output times. To understand this figure,
note that if the contributions from the scalar field potential to the
Poisson equation and scalar field equation of motion are negligible (which
turns out to be the case in our simulations) and if all particles (dark
matter and baryons) have the same coupling to the scalar field, then the
right hand sides of Eqs.~(\ref{eq:INTPoisson}, \ref{eq:INTphiEOM}) are
proportional, with a coefficient $2\gamma $, which implies that $u=2\gamma
\Phi _{c}$; thus Eq.~(\ref{eq:INTdpdtcomov}) says that the strengths of
fifth force ($f$) and gravity ($g$) should satisfy $f=2\gamma ^{2}g$.

In our models only a fraction $\Omega _{\mathrm{CDM}}/\Omega _{m}$, where $%
\Omega _{m}=\Omega _{\mathrm{CDM}}+\Omega _{\mathrm{B}}$, of all the
particles are coupled to the scalar field. Thus, assuming dark matter and
baryonic particles are distributed in the same way, we should have
\begin{eqnarray}\label{eq:f2g_analytic}
f &=& 2\gamma^{2}\frac{\Omega_{\mathrm{CDM}}}{\Omega_{m}}g
\end{eqnarray}
because only a fraction $\Omega _{\mathrm{CDM}}/\Omega _{m}$ of
the
particles which produce gravity also produce the fifth force. Eq.~(\ref%
{eq:f2g_analytic}) is plotted in Fig.~\ref{fig:force} as the solid lines.
The dots are the simulation results for $f/g$ for the particles outputted in
Fig.~\ref{fig:particle}. The agreement between the analytical approximation
and numerical solution is remarkably good, which serves as a test of our
Newton-Gauss-Seidel solver of the scalar field equation of motion%
\footnotemark[2]  .

\footnotetext[2]{%
Remember that the Poisson equation on either the domain grid or the
refinement is solved using different methods from those for the scalar field
equation of motion.}

Furthermore, Fig.~\ref{fig:force} also shows some clearly different feature
from the results of \cite{Li:2010}, in which the fifth force is suppressed
by the chameleon mechanism in certain cases (such as at early times and in
high-density regions). Here, the fifth force is not suppressed because the
potential $V(\varphi )$ is negligible, and the approximation Eq.~(\ref%
{eq:f2g_analytic}) works well any\emph{where }all the \emph{time}. This
confirms our earlier claim that the matter power spectrum will be changed at
early times as will be the initial condition for the $N$-body code.

As should evident now, the spatial configuration of the scalar field $%
\varphi $, or equivalently the rescaled scalar field $u$, should closely
follow that of the gravitational potential $\Phi _{c}$, for the same reasons
that led to Eq.~(\ref{eq:f2g_analytic}). Figs.~\ref{fig:potential} and \ref%
{fig:scalar} confirm this: as can be seen there, for all the models and all
the output times, the configurations of $-u$\footnotemark[3]  and $\Phi _{c}$
are indistinguishable.

\footnotetext[3]{%
Note we expect that $u=2\gamma \Phi _{c}$ approximately as explained above,
and $\gamma <0$: the minus sign here ensures that Fig.~\ref{fig:scalar}
follows Fig.~\ref{fig:potential}, rather than compensating it, so that we
have a better visual impression.}

The above figures show that in our coupled scalar field models the
fifth force $f$ (scalar field $u$) closely mimics the gravity $g$
(gravitational potential $\Phi _{c}$), and so the approximation
$f=2\gamma ^{2}g$, used in many other previous simulation works
(\emph{e.g}.~\cite{Maccio:2004}), is a good one.

\subsection{Matter Power Spectrum}

\label{subsect:Pk}

The nonlinear matter power spectrum measured from the output particle
distribution is of more observational interest and it is the theme of this
subsection. The matter power spectrum in the present work is measured using
\texttt{POWMES} \cite{powmes}, which is a publicly available code based on
the Taylor expansion of trigonometric functions and yields Fourier modes
from a number of fast Fourier transforms controlled by the order of the
expansion.

Fig.~\ref{fig:power} displays the fractional changes of the matter power
spectra $P(k)$ due to the scalar field coupling, where $\Delta P\equiv P_{%
\mathrm{scalar}}-P_{\Lambda \mathrm{CDM}}$. For comparison, we have also
shown the linear results as dashed curves. We can see that on large scales
(small $k$) there is an overall agreement between the linear and nonlinear
predictions, which is not surprising since those scales have not experiences
much nonlinear complication. The agreement is very good at early times (the
top panels) when nonlinear effects generally have not taken place on scales $%
k\sim 0.1~h\mathrm{Mpc}^{-1}$; at late times (top curves in bottom panels)
there is a slight mismatch, which is understandable because our simulation
box is not big enough to allow measurement of $P(k)$ below $k=0.1~h\mathrm{%
Mpc}^{-1}$ where nonlinear effects already enter. Indeed, even in the latter
case, we can see the trend that the linear and nonlinear curves merge
towards at small-$k$. The fact that the nonlinear $P(k)$ reduces to the
linear one on large scales is not trivial as the background cosmology in our
$N$-body simulations has also been modified\footnotemark[4]  (as it is in
\texttt{CAMB}, which is used for the linear calculation), and we take this
as another test of our modified \texttt{MLAPM} code.

\footnotetext[4]{%
As mentioned above: the larger $|\gamma |$, the slower the universe expands
(cf.~Fig.~\ref{fig:background}, lower left panel), and thus the more growth
the matter perturbations experience and the larger $P(k)$ becomes.}

The main advantage of the $N$-body simulation is its ability to probe the
nonlinear structure formation and so we are more concerned with the $P(k)$
on smaller scales, which is also plotted in Fig.~\ref{fig:power}. We can see
that on intermediate scales ($k\sim 1~h\mathrm{Mpc}^{-1}$) the nonlinear $%
P(k)$ beats the linear one, but on even smaller scales it falls behind the
linear value again. For the four models in our simulations the enhancements
of the matter power range from negligible to $\sim 100\%$, but much of the
enhancement is due to the modified background expansion and is already seen
in the linear results. More interestingly, the enhancement is more
significant at earlier times. This can been understood as follows: at later
times the contribution of the dark matter density to the Poisson equation
(and thus the gravitational potential) is decreased due to the coupling
function $C(\varphi )$ becoming increasingly less than 1, weakening further
clustering of the (not only dark but also baryonic) matter. Note that such a
trend can also be seen in the linear $P(k)$, by comparing the bottom panels,
although it is not so obvious.

In order to display the bias developed between baryons and dark matter, we
have also plotted the $P(k)$ for baryons and dark matter separately in Fig.~%
\ref{fig:powerbc}. As expected, the $P(k)$ for dark matter is always larger,
signifying a stronger clustering due to the assistance from the fifth force.
The larger $|\gamma |$ is, the stronger the fifth force will be, and the
larger the bias will become.

\subsection{Mass Function}

\label{subsect:mf}

We identify halos in our $N$-body simulations using \texttt{MHF} (\texttt{%
MLAPM} Halo Finder) \cite{mhf}, which is the default halo finder for \texttt{%
MLAPM}. \texttt{MHF} optimally utilizes the refinement structure of the
simulation grids to pin down the regions where potential halos reside and
organize the refinement hierarchy into a tree structure. \texttt{MLAPM}
refines grids according to the particle density on them and so the
boundaries of the refinements are simply isodensity contours. \texttt{MHF}
collects the particles within these isodensity contours (as well as some
particles outside). It then performs the following operations: (i) assuming
spherical symmetry of the halo, calculate the escape velocity $v_{esc}$ at
the position of each particle, (ii) if the velocity of the particle exceeds $%
v_{esc}$ then it does not belong to the virialized halo and is removed.
Steps (i) and (ii) are then iterated until all unbound particles are removed
from the halo or the number of particles in the halo falls below a
pre-defined threshold, which is 20 in our simulations. Note that the removal
of unbound particles is not used in some halo finders using the spherical
overdensity (SO) algorithm, which includes the particles in the halo as long
as they are within the radius of a virial density contrast. Another
advantage of \texttt{MHF} is that it does not require a pre-defined linking
length in finding halos, such as the friend-of-friend procedure.

As explained in detail in \cite{Li:2010}, part of the \texttt{MHF} algorithm
also needs to be modified for the coupled scalar field model, because the
scalar field $\varphi $ behaves as an extra "potential" (which produces the
fifth force) and so the dark matter particles experience a deeper total
"gravitational" potential than in the $\Lambda $CDM paradigm, all other
things being equal. Consequently, the escape velocity for dark matter
particles increases compared with the Newtonian prediction. As the dark
matter particles in the coupled scalar field simulations are typically
faster than in the $\Lambda $CDM simulation, if we underestimate $v_{esc}$
then some particles which should have remained in the halo will be
incorrectly removed by \texttt{MHF}. In \cite{Li:2010} it was found that the
underestimate of the mass function by using default \texttt{MHF} in the
coupled scalar field models could be up to a few percent.

In the models we are considering here, things are even more complicated. In
\cite{Li:2010}, the chameleon mechanism ensures that the scalar field takes
a very tiny value everywhere, all the time (typically $\sqrt{\kappa }\varphi
\lesssim 10^{-5}$) and so the coupling function $C(\varphi )=e^{\gamma \sqrt{%
\kappa }\varphi }\doteq 1$ is a very good approximation -- this means that
the contribution of the dark matter to the Poisson equation is not
significantly modified compared with that in $\Lambda $CDM. In the models
here, however, $C(\varphi )$ can be up to $30\%$ less than $1$, so that the
source term of the Poisson equation is significantly different from $\Lambda
$CDM.

Obviously, we need to take both of these two major changes into account when
we design a new algorithm to identify halos. An exact analytical calculation
of the escape velocity and Poisson source term in the coupled scalar field
model turns out be difficult, and so we introduce an approximate
alternative, which is based on the \texttt{MHF} default method \cite{ahf}.

The default \texttt{MHF} code works out $v_{esc}$ using the Newtonian result
\begin{eqnarray}  \label{eq:ahf_vec2}
v^{2}_{esc} &=& 2|\Phi|,
\end{eqnarray}
in which $\Phi$ is the gravitational potential. Under the assumption of
\emph{spherical symmetry} for the halos, the Poisson equation $%
\nabla^{2}\Phi=4\pi G\rho_{m}$ could be integrated once to give
\begin{eqnarray}  \label{eq:ahf_newton_law}
\frac{d\Phi}{dr} &=& \frac{GM(<r)}{r^{2}}
\end{eqnarray}
which is just the Newtonian force law. This equation can be integrated once
again to obtain
\begin{eqnarray}  \label{eq:ahf_phi}
\Phi(r) &=& G\int^{r}_{0}\frac{M(<r^{\prime })}{r^{\prime 2}}dr^{\prime }+
\Phi_{0}
\end{eqnarray}
where $\Phi_{0}$ is an integration constant and can be fixed \cite{ahf} by
requiring that $\Phi(r=\infty)=0$ as
\begin{eqnarray}  \label{eq:ahf_phi0}
\Phi_{0} &=& \frac{GM_{vir}}{R_{vir}} + G\int^{R_{vir}}_{0}\frac{%
M(<r^{\prime })}{r^{\prime 2}}dr^{\prime },
\end{eqnarray}
in which $R_{vir}$ is the virial radius of the halo and $M_{vir}$ is the
mass enclosed in $R_{vir}$.

In the coupled scalar field models here, the two modifications mentioned
above are reflected in Eqs.~(\ref{eq:ahf_vec2}, \ref{eq:ahf_newton_law}).

\begin{figure*}[tbp]
\centering \includegraphics[scale=1.9] {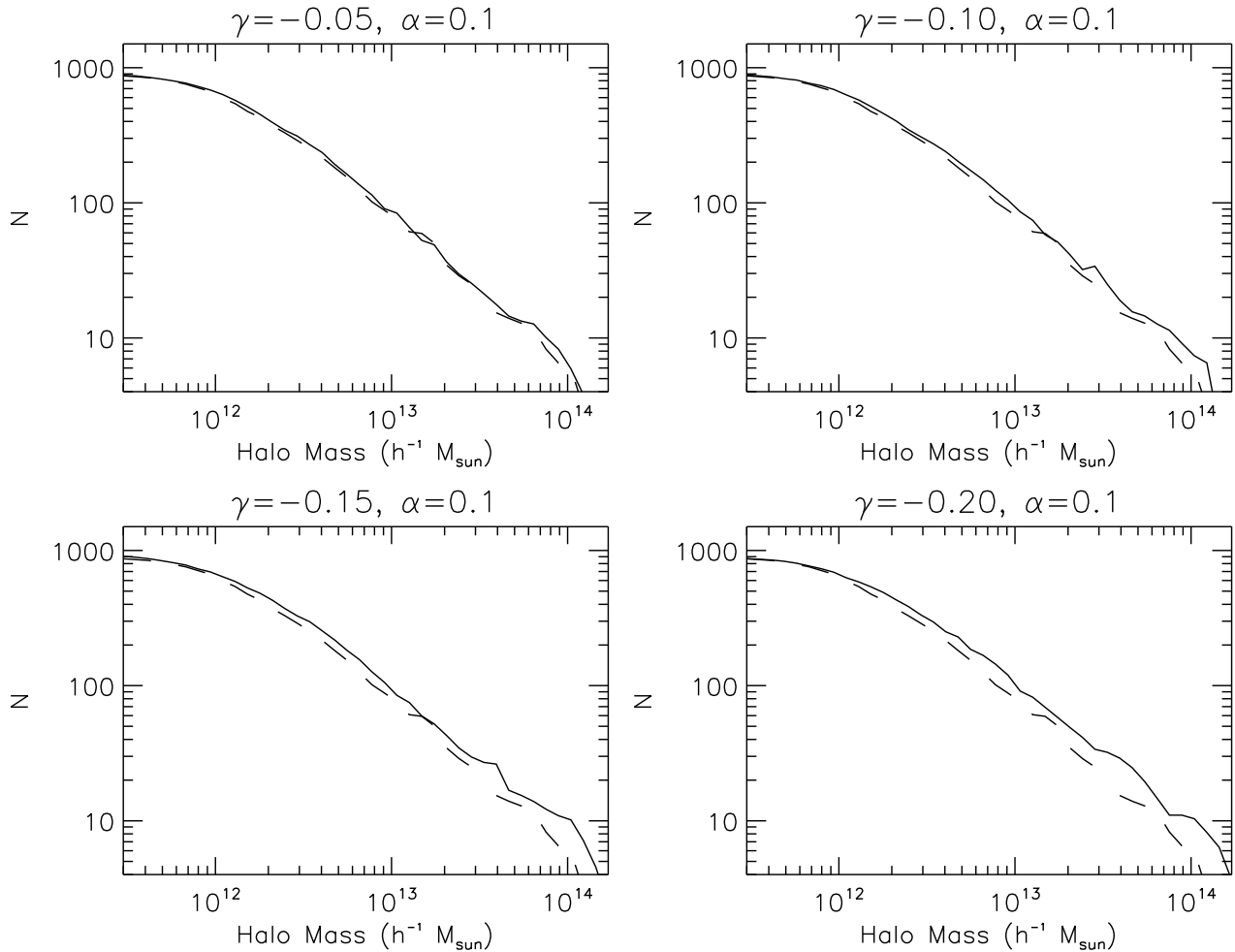}
\caption{The mass functions at $a=1$ for the four coupled scalar field
models under investigation, as indicated by the subtitles of the panels. The
dashed curves are the corresponding result of $\Lambda $CDM.}
\label{fig:mf}
\end{figure*}

\begin{figure*}[tbp]
\centering \includegraphics[scale=1.9] {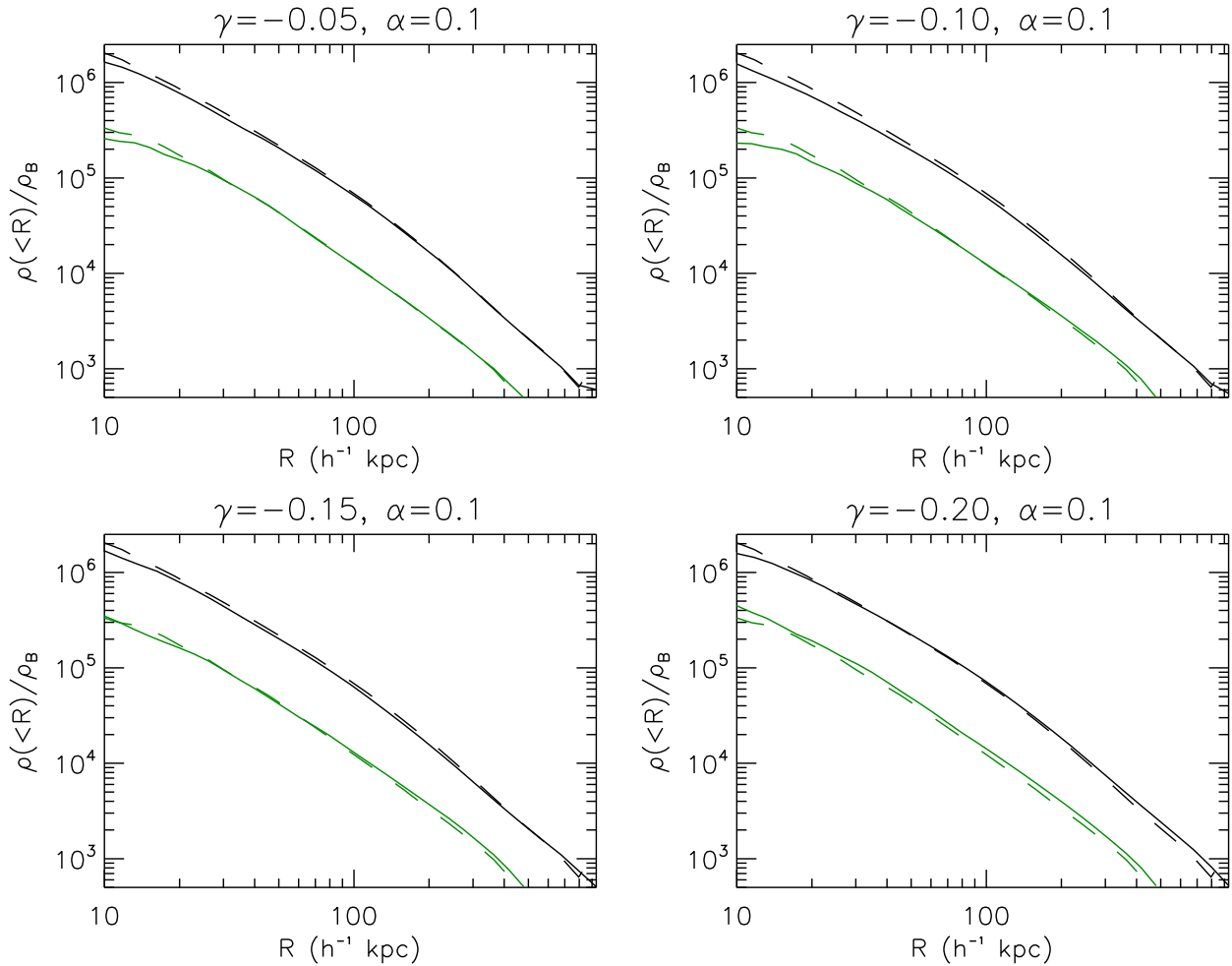}
\caption{(Color Online) The internal density profiled of two selected halos
(see text) at $a=1$. The four panels are for the four models we have
simulated, as illustrated by the subtitles. In all panels solid curves are
for the coupled scalar field model and dashed curves for $\Lambda$CDM. The
upper (black) curves are result for halo I and lower (green) curves for halo
II.}
\label{fig:profile}
\end{figure*}

\begin{figure*}[tbp]
\centering \includegraphics[scale=1.9] {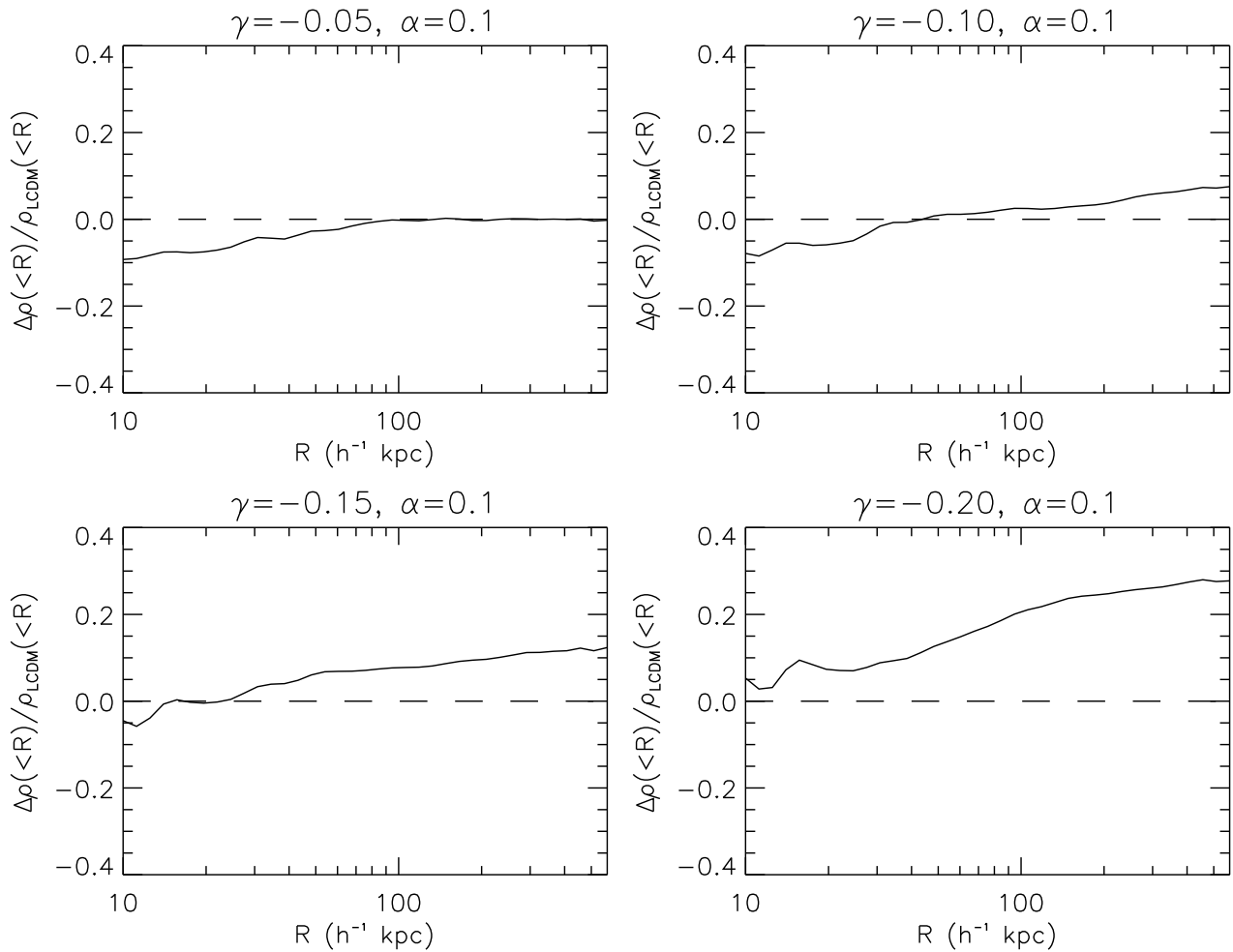}
\caption{The fractional change of the internal density profile due to the
dark matter being coupled to the scalar field, $[\protect\rho(<R)-\protect%
\rho_{\Lambda\mathrm{CDM}}(<R)]/\protect\rho_{\Lambda\mathrm{CDM}}(<R)$. The
solid curves are the average of ten of the heaviest halos in our simulation
boxes and the dashed curve is 0. We show this for the four models we have
simulated, as illustrated by the subtitles.}
\label{fig:profiler}
\end{figure*}

\begin{figure*}[tbp]
\centering \includegraphics[scale=1.9] {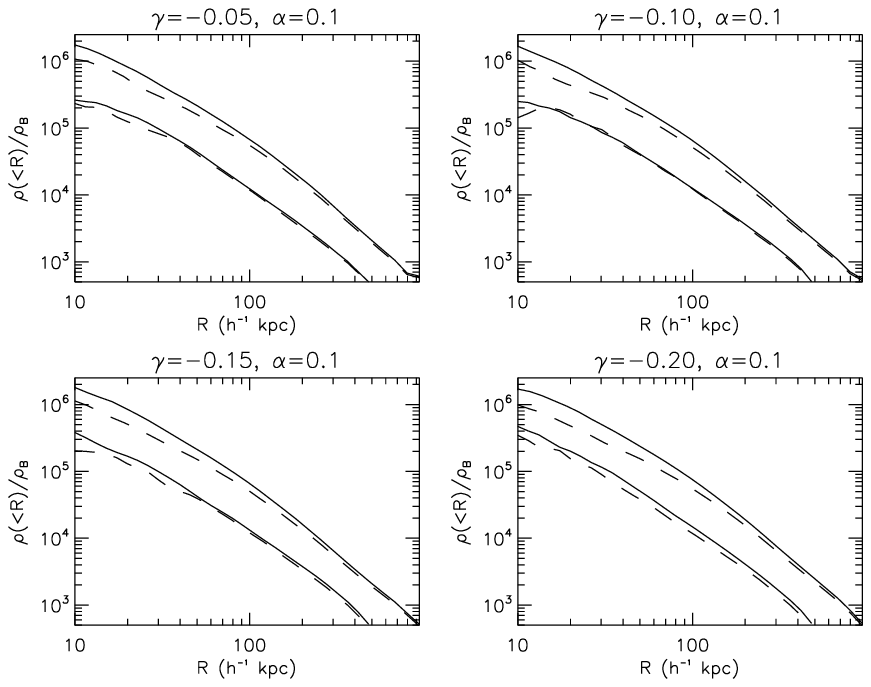}
\caption{The internal density profiles for dark matter (solid curves) and
baryons (dashed curves) separately. We show this for the four models we have
simulated, as indicated by the subtitles. In each panel the upper two curves
are for halo I and the lower two curves for halo II.}
\label{fig:profilebc}
\end{figure*}

For Eq.~(\ref{eq:ahf_vec2}), we realize that the gravitational potential $%
\Phi $ is not the only factor determining the escape velocity: there is also
a contribution from the scalar field $\varphi $. There are different ways to
approach this problem, the simplest of which is to obtain the total
"potential" by rescaling $\Phi $ with a constant (remember we have shown in
Sect.~\ref{subsect:snapshots} that $\varphi \propto \Phi $ is a good
approximation all the time and everywhere). A more delicate approach can be
devised by noting that in the default \texttt{MHF} code, Eq.~(\ref%
{eq:ahf_newton_law}) is used in the numerical integrations to obtain both $%
\Phi (r)$ and $\Phi _{0}$ [cf.~Eqs.(\ref{eq:ahf_phi}, \ref{eq:ahf_phi0})].
More explicitly, the code loops over all particles in the halo in ascending
order of the distance from the halo centre, and whenever a particle is
encountered its mass is uniformly distributed into the spherical shell
between the particle and its previous particle (the thickness of the shell
is now the $dr$ of the integration). When the fifth force is added, we call
its contribution to the total "gravitational" potential $\Phi _{\varphi }$
and its value at infinity, $\Phi _{\varphi 0}$. Our estimate of the escape
velocity is now, instead of Eq.~(\ref{eq:ahf_vec2}), given by
\begin{equation}\label{eq:ahf_vescnew}
v_{esc}^{2}=2|\Phi +\Phi _{\varphi }-\Phi _{0}-\Phi _{\varphi 0}|.
\end{equation}

We have recorded the components of gravity and the fifth force for each
particle in the simulation, and so the ratio $f/g$ can be computed at the
position of each particle, which gives $\left( \Phi _{\varphi }-\Phi
_{\varphi 0}\right) /\left( \Phi -\Phi _{0}\right) $ at that position, which
can be used in Eq.~(\ref{eq:ahf_vescnew}). In this way we have, at least
approximately, taken into account the fifth force and $\Phi _{\varphi }$.

For Eq.~(\ref{eq:ahf_newton_law}), which is used to compute $\Phi $ and $%
\Phi _{0}$, we know that it is applicable in the coupled scalar field model
as well, because the contribution from the scalar field potential is
negligible. But one should be careful when interpreting the mass $M$ here:
it is \emph{not} the total bare mass of all the particles ($M_{\mathrm{B}%
}+M_{\mathrm{CDM}}$) within radius $r$, but $M_{\mathrm{B}}+C(\varphi )M_{%
\mathrm{CDM}}$. Such a fact can be taken into account by multiplying the
mass of each \emph{dark matter} (not baryonic) particle by $C(\varphi ),$
which is evaluated using the $\varphi $ at the position of that particle
(which we have recorded too). Note that observations of the masses are made
by registering gravitational effect, and so the halo mass we measure should
be $M_{\mathrm{B}}+C(\varphi )M_{\mathrm{CDM}}$ rather than $M_{\mathrm{B}%
}+M_{\mathrm{CDM}}$: in what follows we will always talk about the former
unless otherwise stated.

In Fig.~\ref{fig:mf} we have plotted the mass functions of our four coupled
scalar field models at the current time, as compared with the prediction of
the $\Lambda $CDM paradigm. Obviously, as the coupling between dark matter
and the scalar field becomes stronger, more heavy halos will be produced
during structure formation. However, it is not yet clear which physical
effect is mainly responsible for such a pattern, and several factors could
be crucial. For examples, as $|\gamma |$ increases, the fifth force
strengthens and the background expansion gets slower, both of which would
help boost the clustering of matter; on the other hand, the dark matter's
contribution to the Poisson equation is weakened due to $C(\varphi )$
becoming increasingly less than $1$ -- this will weaken the clustering of
matter. Detailed study of the significance of these different physical
effects is beyond the scope of this work.

\subsection{Halo Properties}

\label{subsect:halo}

Finally, we are also interested in the properties of the dark matter halos
in the coupled scalar field models. In the case of a chameleon-like scalar
field, ref \cite{Li:2010} shows that there is a strong environmental
dependence on density profile and baryon-dark-matter bias inside the halo,
which is controlled by the model parameters in a complicated way. In the
models generated here, the fifth force is everywhere unsuppressed and we do
not expect such an environmental dependence as seen in \cite{Li:2010}.
However, we have several factors mentioned in Sect.~\ref{subsect:mf}, which
can potentially lead to interesting new features.

We will look first at the internal density profiles, which can be measured
down to small radii thanks to the higher spatial resolution of the
self-refinement code. We have selected two typical halos from each
simulation. Halo I is centred on $(x,y,z)\sim (32.4,31.5,61.2)h^{-1}$~Mpc,
which is slightly different for different simulations, and has a virial mass
$M_{vir}\sim 1.29\times 10^{14}h^{-1}~M_{\bigodot }$; halo II is centred on $%
(x,y,z)\sim (54.8,39.8,35.7)h^{-1}$~Mpc, which is also slightly different
for different simulations, and has a virial mass $M_{vir}\sim 1.90\times
10^{13}h^{-1}~M_{\bigodot }$ (note here that the virial masses are for the
\emph{$\Lambda $CDM simulations}; for scalar-field simulations they can be,
and generally are, larger).

Fig.~\ref{fig:profile} summarizes what we have, and it shows that the
density profile for the coupled scalar field model is rather similar to that
of $\Lambda $CDM, except that for the innermost part the overdensity for the
scalar field model (solid curves) drops below that of $\Lambda $CDM (dashed
curves). Such a phenomenon has previously been detected by the authors of
\cite{Baldi:2008} and explained as mainly due to the velocity-dependent
acceleration of dark matter particles [the last term in Eq.~(\ref%
{eq:DMEOM_physical})]. However, here we find here that such a
trend does \emph{not} continue with increasing $|\gamma |$ and is
indeed reversed for large $|\gamma |$ values. To see this more
explicitly, in Fig.~\ref{fig:profiler} we have shown the quantity
$\Delta \rho (<R)/\rho _{\Lambda \mathrm{CDM}}(<R)$ (where $\rho
(<R)$ is the average density inside radius $R$ and $\Delta $ means
the difference between the coupled scalar field model and $\Lambda
$CDM) averaged over 10 of the heaviest halos of our simulations.
This shows clearly that for small $|\gamma |$ values the inner
density is systematically lower in our coupled scalar field models
than in $\Lambda $CDM while this is not necessarily true for the
density in the outer part of the halos. Also, as $|\gamma |$
increases, the lowering of inner density becomes less significant,
and finally for $\gamma =-0.20$ we see the opposite trend emerge.
The physical explanations of the observed patterns are not of
prime interest to our study here.

Next, we can look at the bias between baryons and dark matter in the halos,
which is displayed in Fig.~\ref{fig:profilebc}. As expected, we find that
the dark matter density is constantly higher than that of baryons, due to
the extra force they feel (note that other factors, such as the cosmic
expansion rate and the modification of the source term in the Poisson
equation, have the same influence on dark matter and baryons). In general,
the bias increases with the coupling strength $|\gamma |$ which is easy to
understand.

\section{Summary and Discussion}

\label{sect:con}

Couplings between any scalar field and (some of) the matter species can
affect cosmology in various ways. There are two principal effects: modified
source terms in the gravitational field equations (\emph{channel I}) and
direct new interactions between particles of the coupled matter species (%
\emph{channel II}).

The effects through channel II arise in an inhomogeneous universe, but not
in background cosmology. In principle, there is only effect of this sort --
the fifth force, due to the exchanges of scalar quanta between matter
particles. In practice, (\emph{e.g.}, in $N$-body simulations), the scalar
field $\varphi $ is computed in the fundamental observer's frame while the
fifth force must be evaluated in individual particle frames; consequently,
the fifth force is split into two parts: the part from the spatial gradient
of the scalar field $\delta \varphi $ in the fundamental observer's frame
(which is independent of the particle's peculiar velocity $v$) and the part
due to frame transformation which is proportional to $v$. In a homogeneous
universe, there is no spatial gradient and all matter particles are comoving
with the fundamental observers, so $v=0$, and therefore the channel II
effects vanish. In an inhomogeneous universe both terms are nonzero in
general, leading to a net force which strengthens the attraction between
particles, and enhances their gravitational clustering.

The effects arising through channel I appear both in homogeneous (via the
Friedmann and Raychaudhuri equations) and inhomogeneous (via the Poisson
equation) universes. They arise mainly via a renormalization of the
contributions to the total density from the coupled matter species and by a
new contribution from the scalar field itself. \ The exact effects depend on
the specific forms of the scalar potential and coupling function. Here, we
list our main findings for an inverse power-law potential Eq.~(\ref%
{eq:potential}), with an exponential coupling Eq.~(\ref{eq:coupling_function}%
), and the model parameters given in Sect.~\ref{subsect:param}.

\footnotetext{%
Note in Fig.~\ref{fig:background} (lower left panel) the expansion rate of
the coupled scalar field model finally catches up with that of $\Lambda $CDM
at $a=1$. This is because at late times the dark energy (scalar field)
density is higher than that in $\Lambda $CDM.}

For the background cosmology, the contribution to the Friedmann equation
from dark matter is renormalized by $C(\varphi )<1$, and $C(\varphi )$
decreases as the coupling constant $|\gamma |$ increases. As a result, a
stronger coupling (larger $|\gamma |$) produces lower cosmic expansion rate
during the matter-dominated era (Fig.~\ref{fig:background})\footnotemark[5]
, which will in turn helps enhance the clustering of matter and promotes
structure formation.

Such an enhancement is just what we have observed from our linear
perturbation analysis (Figs.~\ref{fig:linpert} and \ref{fig:linpert2}): on
very large scales, which are beyond the range of the fifth force effects,
the matter power spectrum increases with $|\gamma |$. On small scales, the
fifth force helps to enhance the matter power, making it increase further
than predicted in $\Lambda $CDM (comparing the left to the right panels of
Fig.~\ref{fig:linpert2}, where we have separated baryons, which have no
scalar coupling, from dark matter, which has). This enhancement starts at a
rather early, and so the scalar field does leave imprints on the matter
power spectrum at $z\sim 50$, which means that the initial condition for $N$%
-body simulations also needs to modified. For example, in the model with $%
| \gamma |=0.15$, we have found that :

(i) The density perturbation at $z\sim 49$ is about $10\%$ larger than the
corresponding result in $\Lambda $CDM for baryons (due to slower cosmic
expansion) and about $12\%$ larger for dark matter (due to the slower
expansion \emph{and} the fifth force effects);

(ii) The average velocity at the same time is about $8\%$ larger than the
corresponding result in $\Lambda $CDM for baryons and about $10\%$ larger
for dark matter. We introduce a quick method to take these changes into
account when generating initial conditions for $N$-body simulations.

One of the most important results from our $N$-body simulations is the
magnitude of fifth force. Fig.~\ref{fig:force} shows that everywhere and at
any time (\emph{i.e.}, in both high and low density regions) the fifth force
(or more precisely, the velocity-\emph{in}dependent part of it) is
proportional to gravity in magnitude (with a coefficient $2\gamma ^{2}$ when
only considering dark matter). This is the approximation used in many
previous simplified simulations, and our results confirm numerically that
this approximation works fairly well, at least for those cases where the
scalar potential is unimportant, such as ours. We emphasize, however, that
the velocity-dependent part of the fifth force should \emph{not} be dropped
(as in some previous simulations) unless there is good reason to do so (see
\cite{Li:2010}).

For the nonlinear matter power spectrum (Fig.~\ref{fig:power}), we find that
the $N$-body simulations show agreement with linear perturbation analysis on
large scales. On intermediate scales, our simulations predict significant
enhancement in the matter power over that predicted both by linear
perturbation analysis and by the $\Lambda $CDM paradigm. This enhancement is
strongest at early epochs but gradually weakens at late times. One possible
reason for this is that the scalar coupling, $C(\varphi ),$ decreases with
time, and reduces the contribution of dark matter to the gravitational
potential, so weakening further clustering, but this still needs to be
investigated in more detail.

The bias between dark matter and baryons increases with time and with
coupling strength $|\gamma |$. This is because larger $|\gamma |$ implies
stronger fifth forces and therefore stronger total forces act on dark matter
particles compared to baryons. As time passes, the bias has more time to
develop and so increases as well (Fig.~\ref{fig:powerbc}). Using the same
reasoning, the fifth force increases the attraction of dark matter particles
and so heavier halos are expected to form during the structure formation
(Fig.~\ref{fig:mf}). In order to identify gravitationally bound and
virialized halos in our models we must also take into account this scalar
coupling (Sect.~\ref{subsect:halo}).

The scalar field coupling also has impacts on the internal density profiles
of the halos (Figs.~\ref{fig:profile} and \ref{fig:profiler}). We average
ten of the heaviest halos and find that

(i) When compared with $\Lambda $CDM, the overdensities in the inner regions
can be lower while those in the outer regions is higher, showing the failure
of the halos to retain particles in the inner region, either because the
particles move too fast or because the attractive potential at the centre is
too weak;

(ii) The suppression of inner density compared with $\Lambda $CDM weakens,
rather than strengthens, as $|\gamma |$ increases. Indeed, for $|\gamma
| =0.2 $ the density is higher than $\Lambda $CDM result throughout the
halos. More detailed studies are needed to clarify the leading effect
responsible for this observed pattern.

(iii) There is a bias between density profiles for baryons and dark matter
(Fig.~\ref{fig:profilebc}): the dark matter density is always higher because
it experiences the fifth force which boosts its clustering.

In summary, in this paper we have been given a comprehensive description of
the methodology and implementation of general $N$-body simulations for
coupled scalar field models. Some important issues in these simulations have
been addressed here for the first time: the consistent solution of the
scalar field and fifth force, the fifth force effects on generating initial
conditions, and the effects of scalar field on identifying virialized halos.
Although the situation is complex, we have identified interesting new
features in these models. We hope these developments will lead to more
detailed study of nonlinear structure formation in this class of models and
facilitate their subsequent confrontation with observational data.

\begin{acknowledgments}
The work described in this paper has been performed on \texttt{COSMOS}, the
UK National Cosmology Supercomputer. The coding work at early stage was done
on the \texttt{SARA} Supercomputer in the Netherlands, under the \texttt{%
HPC-EUROPA2} project, with the support of the European Community
Research Infrastructure Action under the FP8 Programme. The linear
perturbation calculations in this work are performed using
\texttt{CAMB} \cite{Lewis:2000}. We thank Luca Amendola, Marco
Baldi, Kazuya Koyama, Andrea Maccio, Gong-Bo Zhao and Hongsheng
Zhao for useful conversations relevant to this work, and computing
staff of DAMTP for technical help in the usage of \texttt{COSMOS}.
B.~Li is supported by the Research Fellowship at Queens' College,
Cambridge, and the STFC.
\end{acknowledgments}

\bigskip

\appendix

\section{Discretized Equations for the $N$-body Simulations}

\label{appen:discret}

In the \texttt{MLAPM} code for Poisson equation Eq.~(\ref{eq:INTPoisson}) is
(and in our modified code the scalar field equation of motion Eq.~(\ref%
{eq:INTphiEOM}) will also be) solved on discretized grid points, so we must
develop the discrete versions of Eqs.~(\ref{eq:INTdxdtcomov} - \ref%
{eq:INTphiEOM}) to be implemented in the code. First, we write down the full
formalism of the relevant equations.

Introducing the variable $u$ (cf.~Sect.~\ref{subsect:codeunit}), the Poisson
equation becomes
\begin{eqnarray}\label{eq:u_Poisson}
&&\nabla^{2}\Phi_{c}\nonumber\\ &=&
\frac{3}{2}\Omega_{\mathrm{CDM}}\left\{\rho_{c,\mathrm{CDM}}\exp\left[\gamma\left(\sqrt{\kappa}\bar{\varphi}+\frac{B^2H^2_0}{ac^{2}}u\right)\right]
-
e^{\gamma\sqrt{\kappa}\bar{\varphi}}\right\}\nonumber\\
&& +
\frac{3}{2}\Omega_{\mathrm{B}}\left(\rho_{c,\mathrm{B}}-1\right) -
\frac{3\lambda
a^{3}}{\left[\sqrt{\kappa}\bar{\varphi}+\frac{B^2H^2_0}{ac^{2}}u\right]^{\alpha}}
+ \frac{3\lambda
a^{3}}{\left[\sqrt{\kappa}\bar{\varphi}\right]^{\alpha}},\ \ \ \ \
\end{eqnarray}
where $\lambda $ is defined in Eq.~(\ref{eq:newphi}) and is a constant of $%
O(1)$ (actually $\lambda \sim \Omega _{\mathrm{DE}}$ where $_{\mathrm{DE}}$
means dark energy). Its value and that of the quantity $\sqrt{\kappa }\bar{%
\varphi}$ are determined solely by the background evolution. The computation
of the background quantities is discussed in Sect.~\ref{subsect:bkgd}.

The scalar field equation of motion now becomes
\begin{eqnarray}\label{eq:u_phi_EOM}
\nabla^2 u &=& 3\gamma\Omega_{\mathrm{CDM}}\rho_{c,\mathrm{CDM}}
\exp\left[\gamma\left(\sqrt{\kappa}\bar{\varphi}+\frac{B^2H^2_0}{ac^{2}}u\right)\right]\nonumber\\
&& - \frac{3\alpha\lambda a^{3}}
{\left[\sqrt{\kappa}\bar{\varphi}+\frac{B^2H^2_0}{ac^{2}}u\right]^{1+\alpha}}
- 3\gamma\Omega_{\mathrm{CDM}}e^{\gamma\sqrt{\kappa}\bar{\varphi}}\nonumber\\
&& + \frac{3\alpha\lambda a^{3}}
{\left[\sqrt{\kappa}\bar{\varphi}\right]^{1+\alpha}}.
\end{eqnarray}

So, in terms of the new variable $u$, the set of equations used in the $N$%
-body code is Eqs.~(\ref{eq:INTdxdtcomov}, \ref{eq:INTdpdtcomov}) plus Eqs.~(%
\ref{eq:u_Poisson}, \ref{eq:u_phi_EOM}). These equations were used in the
code. Among them, Eqs.~(\ref{eq:u_Poisson}, \ref{eq:INTdpdtcomov}) will use
the value of $u,$ while Eq.~(\ref{eq:u_phi_EOM}) solves for $u$. In order
that these equations can be integrated into \texttt{MLAPM}, we need to
discretize Eq.~(\ref{eq:u_phi_EOM}) for the application of
Newton-Gauss-Seidel relaxation method. This involves writing down a discrete
version of this equation on a uniform grid with grid spacing $h$. Suppose we
want to achieve second-order precision, as is in the default Poisson solver
of \texttt{MLAPM}, then $\nabla ^{2}u$ in one dimension can be written as
\begin{eqnarray}
\nabla^{2}u &\rightarrow& \nabla^{h2}u_{j}\ =\
\frac{u_{j+1}+u_{j-1}-2u_{j}}{h^2}
\end{eqnarray}
where a subscript $_{j}$ means that the quantity is evaluated on
the $j$-th point. The generalization to three dimensions is
straightforward.

The discrete version of Eq.~(\ref{eq:u_phi_EOM}) is
\begin{eqnarray}\label{eq:diffop}
L^{h}\left(u_{i,j,k}\right) &=& 0,
\end{eqnarray}
in which
\begin{widetext}
\begin{eqnarray}
L^{h}\left(u_{i,j,k}\right) &=&
\frac{1}{h^{2}}\left[u_{i+1,j,k}+u_{i-1,j,k}+u_{i,j+1,k}+u_{i,j-1,k}+u_{i,j,k+1}+u_{i,j,k-1}-6u_{i,j,k}\right]\nonumber\\
&&-\left[3\gamma\Omega_{\mathrm{CDM}}\rho_{c,\mathrm{CDM}}
\exp\left[\gamma\left(\sqrt{\kappa}\bar{\varphi}+\frac{B^2H^2_0}{ac^{2}}u_{i,j,k}\right)\right]
- \frac{3\alpha\lambda a^{3}}
{\left[\sqrt{\kappa}\bar{\varphi}+\frac{B^2H^2_0}{ac^{2}}u_{i,j,k}\right]^{1+\alpha}}\right]\nonumber\\
&&
+\left[3\gamma\Omega_{\mathrm{CDM}}e^{\gamma\sqrt{\kappa}\bar{\varphi}}
-\frac{3\alpha\lambda a^{3}}
{\left[\sqrt{\kappa}\bar{\varphi}\right]^{1+\alpha}}\right].
\end{eqnarray}
\end{widetext}
Then, the Newton-Gauss-Seidel iteration says that we can
obtain a new (and often more accurate) solution of $u$, $u_{i,j,k}^{\mathrm{%
new}}$, using our knowledge about the old (and less accurate) solution $%
u_{i,j,k}^{\mathrm{old}}$ via
\begin{eqnarray}\label{eq:GS}
u^{\mathrm{new}}_{i,j,k} &=& u^{\mathrm{old}}_{i,j,k} -
\frac{L^{h}\left(u^{\mathrm{old}}_{i,j,k}\right)}{\partial
L^{h}\left(u^{\mathrm{old}}_{i,j,k}\right)/\partial u_{i,j,k}}.
\end{eqnarray}
The old solution will be replaced by the new solution to
$u_{i,j,k}$ once the new solution is ready, using the red-black
Gauss-Seidel sweeping scheme. Note that
\begin{widetext}
\begin{eqnarray}
\frac{\partial L^{h}(u_{i,j,k})}{\partial u_{i,j,k}} &=&
-\frac{6}{h^{2}}
-3\gamma^{2}\frac{\left(H_{0}B\right)^{2}}{ac^{2}}\Omega_{\mathrm{CDM}}\rho^{\mathrm{CDM}}_{c,i,j,k}
\exp\left[\gamma\left(\sqrt{\kappa}\bar{\varphi}+\frac{B^2H^2_0}{ac^{2}}u_{i,j,k}\right)\right]
-\frac{\left(H_{0}B\right)^{2}}{ac^{2}}\frac{3\alpha(1+\alpha)\lambda
a^{3}}
{\left[\sqrt{\kappa}\bar{\varphi}+\frac{B^2H^2_0}{ac^{2}}u_{i,j,k}\right]^{2+\alpha}}.\
\ \
\end{eqnarray}
\end{widetext}

In principle, if we start from a high redshift, then the initial guess of $%
u_{i,j,k}$ for the relaxation could be so chosen that the initial value of $%
\varphi $ in all space is equal to the background value $\bar{\varphi}$,
because at this time we expect this to be approximately true any way. At
subsequent time-steps we could use the solution for $u_{i,j,k}$ from the
previous time-step as our initial guess. If the timestep is small enough
then we would expect $u$ to change only slightly between consecutive
timesteps so that such a guess will be good enough for the iterations to
converge quickly.

In practice, however, due to specific features and the algorithm of the
\texttt{MLAPM} code \cite{Knebe:2001}, the above procedure may be slightly
different in details.

\section{The Dark Matter Lagrangian}

\label{appen:Lagrangian}

For simplicity let us write the CDM Lagrangian as $\mathcal{L}_{\mathrm{CDM}%
}=F^{1/2}=\sqrt{g_{ab}\dot{x}^{a}\dot{x}^{b}}$ (\emph{i.e.}, neglecting the
mass term and $\delta $-function term in Eq.~(\ref{eq:DMLagrangian})) in
which $\dot{x}^{a}\equiv dx^{a}/ds$ and $s$ is some general parameterization
of a particle's geodesic (we only use this definition in this appendix and
note that in other places in the paper $\dot{x}$ has different meaning). The
Euler-Lagrange equation is
\begin{eqnarray}
\frac{d}{ds}\frac{\partial
\mathcal{L}\mathrm{CDM}}{\partial\dot{x}^{a}} &=& \frac{\partial
\mathcal{L}_{\mathrm{CDM}}}{\partial x^{a}}
\end{eqnarray}
in which we have
\begin{eqnarray}
\frac{\partial \mathcal{L}_{\mathrm{CDM}}}{\partial x^{a}} &=&
\frac{1}{2}F^{-1/2}g_{bc,a}\dot{x}^{b}\dot{x}^{c},\nonumber\\
\frac{\partial \mathcal{L}_{\mathrm{CDM}}}{\partial\dot{x}^{a}}
&=&
\frac{1}{2}F^{-1/2}\frac{\partial}{\partial\dot{x}^{a}}(g_{bc}\dot{x}^{b}\dot{x}^{c})\nonumber\\
&=& \frac{1}{2}F^{-1/2}g_{bc}(\delta^{b}_{a}\dot{x}^{c} +
\delta^{c}_{a}\dot{x}^{b}),\nonumber
\end{eqnarray}
and
\begin{eqnarray}
&&\frac{d}{ds}\frac{\partial
\mathcal{L}_{\mathrm{CDM}}}{\partial\dot{x}^{a}}\nonumber\\ &=&
F^{-1/2}\left[g_{ab}\ddot{x}^{b} -
\frac{1}{2}F^{-1}\dot{F}g_{ab}\dot{x}^{b} +
\frac{1}{2}(g_{ac,b}\dot{x}^{b}\dot{x}^{c} +
g_{ab,c}\dot{x}^{b}\dot{x}^{c})\right]\nonumber
\end{eqnarray}
so that the Lagrange equation becomes
\begin{eqnarray}\label{eq:appen_DMLagrangian1}
\ddot{x}^{d} - \frac{1}{2}F^{-1}\dot{F}\dot{x}^{d} +
\Gamma^{d}_{bc}\dot{x}^{b}\dot{x}^{c} &=& 0.
\end{eqnarray}
For timelike geodesics we can choose $s=\tau $ ($\tau $ being the proper
time) which means that $F=1$ so that Eq.~(\ref{eq:appen_DMLagrangian1})
becomes the usual geodesic equation
\begin{eqnarray}\label{eq:appen_DMLagrangian2}
\ddot{x}^{a} + \Gamma^{a}_{bc}\dot{x}^{b}\dot{x}^{c} &=& 0.
\end{eqnarray}
If $F\neq 1,$ then we need to retain Eq.~(\ref{eq:appen_DMLagrangian1}) as
the geodesic equation, which is unfamiliar. We note that $F=1$ (and $%
\mathcal{L}_{\mathrm{CDM}}=1$ above) corresponds to the result in our Eq.~(%
\ref{eq:DMLagrangian2}). As it stands, Eq.~(\ref{eq:DMLagrangian2}) is
exact. Since we use proper time $\tau$ for $s$, the geodesic equation is
also written in terms of $\tau$, which means that when we switch from $\tau
$ to the physical time $t$ in Eq.~(\ref{eq:DMEOM}) there will be some
approximation, but Eq.~(\ref{eq:DMEOM}) is approximate in any case.

\section{An Algorithm to Solve for the Background Evolution}

\label{appen:bkgd}

Here, we aim to give the formulae and an algorithm for the background field
equations which can also be applied to linear Boltzmann codes such as
\texttt{CAMB}. Throughout this Appendix we use the conformal time $\tau $
instead of the physical time $t$, and $^{\prime }\equiv d/d\tau ,\mathcal{H}%
\equiv a^{\prime }/a$. All quantities appearing here are background ones
unless stated otherwise.

For convenience, we will work with dimensionless quantities and define $%
\tilde{\varphi}\equiv \sqrt{\kappa }\varphi $ and $N\equiv \ln a$ so that
\begin{eqnarray}
\tilde{\varphi}' &=& \mathcal{H}\frac{d\tilde{\varphi}}{dN},\\
\tilde{\varphi}'' &=&
\mathcal{H}^{2}\frac{d^2\tilde{\varphi}}{dN^2} +
\mathcal{H}'\frac{d\tilde{\varphi}}{dN}.
\end{eqnarray}
With these definitions it is straightforward to show that Eq.~(\ref%
{eq:bkgd_scalar}) can be expressed as
\begin{eqnarray}\label{eq:bkgd_scalar2}
\left(\frac{\mathcal{H}}{\mathcal{H}_{0}}\right)^2\frac{d^{2}\tilde{\varphi}}{dN^{2}}
+
\left(2\frac{\mathcal{H}^2}{\mathcal{H}_0^{2}}+\frac{\mathcal{H}'}{\mathcal{H}_0^{2}}\right)\frac{d\tilde{\varphi}}{dN}\nonumber\\
-3\alpha\lambda a^{2}\frac{1}{\tilde{\varphi}^{1+\alpha}} +
\frac{3}{a}\gamma\Omega_{\mathrm{CDM}}e^{\gamma\tilde{\varphi}}
&=& 0,
\end{eqnarray}
where $\mathcal{H}_{0}$ is the current value of $\mathcal{H}$, and
the
coefficients of the derivative terms are given by Eqs.~(\ref%
{eq:bkgd_friedman}, \ref{eq:bkgd_raychaudhuri}) as
\begin{eqnarray}\label{eq:bkgd_friedman2}
&&\left(\frac{\mathcal{H}}{\mathcal{H}_0}\right)^2\nonumber\\ &=&
\frac{\Omega_{\mathrm{RAD}}a^{-2}+\Omega_{\mathrm{B}}a^{-1}+e^{\gamma\tilde{\varphi}}
\Omega_{\mathrm{CDM}}a^{-1}+\frac{\lambda
a^{2}}{\tilde{\varphi}^{\alpha}}}{1-\frac{1}{6}\left(\frac{d\tilde{\varphi}}{dN}\right)^2},\
\
\end{eqnarray}
and
\begin{eqnarray}\label{eq:bkgd_raychaudhuri2}
\frac{\mathcal{H}'}{\mathcal{H}^2_0} &=&
-\frac{1}{3}\left(\frac{d\tilde{\varphi}}{dN}\right)^2\frac{\mathcal{H}^{2}}{\mathcal{H}_0^2}
+ \frac{\lambda a^{2}}{\tilde{\varphi}^{\alpha}}\nonumber\\
&&-\left[\Omega_{\mathrm{RAD}}a^{-2}+\frac{1}{2}\Omega_{\mathrm{B}}a^{-1}+\frac{1}{2}e^{\gamma\tilde{\varphi}}
\Omega_{\mathrm{CDM}}a^{-1}\right].\ \ \
\end{eqnarray}
In all of the above equations we have used the facts that
\begin{eqnarray}
\rho_{\mathrm{CDM}} &\propto& a^{-3},\nonumber\\
\rho_{\mathrm{B}} &\propto& a^{-3},\nonumber\\
\rho_{\mathrm{RAD}} &\propto& a^{-4}.\nonumber
\end{eqnarray}
We stress again that although coupled to the scalar field, the
background
dark-matter energy density follows the same conservation law as in $\Lambda $%
CDM. The effect of the coupling is reflected in the fact that there is a
coefficient $e^{\gamma \tilde{\varphi}}$ in front of $\Omega _{\mathrm{CDM}}$
whenever the latter appears in the gravitational equations or the scalar
field evolution equation \cite{Li:2009sy}.

When solving for $\varphi $ (or $\tilde{\varphi}$), we just use Eq.~(\ref%
{eq:bkgd_scalar2}) aided by Eqs.~(\ref{eq:bkgd_friedman2}, \ref%
{eq:bkgd_raychaudhuri2}). It may appear then that, given any initial values
for $\tilde{\varphi}_{\mathrm{ini}}$ and $\left( d\tilde{\varphi}/dN\right)
_{\mathrm{ini}},$ the evolution of $\varphi $ is obtainable. However, Eq.~(%
\ref{eq:bkgd_friedman2}) is not necessarily satisfied for $\tilde{\varphi}$
evolved in such way. Instead, it constrains the initial condition $\tilde{%
\varphi}$ must start with, and the way it must subsequently evolve. This in
turn is determined by the parameters $\lambda ,\alpha ,\gamma $; since $%
\alpha ,\gamma $ specify a model and are fixed once the model is chosen, the
only concern is $\lambda $.

For the initial conditions $\tilde{\varphi}_{\mathrm{ini}}$ and $\left( d%
\tilde{\varphi}/dN\right) _{\mathrm{ini}}$, we have found that the
subsequent evolution of $\tilde{\varphi}$ is rather insensitive to them.
Thus, we choose $\tilde{\varphi}_{\mathrm{ini}}=\left( d\tilde{\varphi}%
/dN\right) _{\mathrm{ini}}=0$ at some very early time (say $N_{\mathrm{ini}}$
corresponds to $a_{\mathrm{ini}}=e^{N_{\mathrm{ini}}}=10^{-6}$) in all the
models. Such a choice is clearly not only practical but also reasonable,
given the fact that we expect that the scalar field starts high up the
potential and rolls down subsequently.

As for $\lambda $, we use a trial-and-error method to find its value which
ensures that (here a subscript $_{0}$ denotes the present-day value)
\begin{eqnarray}
\Omega_{\mathrm{RAD}}+\Omega_{\mathrm{B}}+e^{\gamma\tilde{\varphi}_{0}}
\Omega_{\mathrm{CDM}}+\frac{\lambda}{\tilde{\varphi}_{0}^{\alpha}}
&=&
1-\frac{1}{6}\left(\frac{d\tilde{\varphi}}{dN}\right)^2_{0}\nonumber
\end{eqnarray}
which comes from setting $a=1$ in Eq.~(\ref{eq:bkgd_friedman2}).

We determine the correct value of $\lambda $ for any given $\alpha ,\gamma $
in this way using \texttt{MAPLE}, and then compute the values of $\tilde{%
\varphi}$ and $d\tilde{\varphi}/dN$ for predefined values of $N$ stored in
an array. Their values at any time are then obtained using interpolation,
and with these it is straightforward to compute other relevant quantities,
such as $\mathcal{H},\mathcal{H}^{\prime },$and $\varphi ,$ which are used
in the Boltzmann and $N$-body codes.

\section{The Zeldovich Approximation}

\label{appen:Zeldovich}

The initial conditions for $N$-body simulations are conventionally generated
using the Zeldovich approximation \cite{Zeldovich, Efstathiou:1985}, which
in its original form works only for non-coupled dark matter. When there is a
non-minimal coupling between dark matter particles and a scalar field, it
must be generalized, and we discuss this here.

Consider a particle whose actual position $\mathbf{r}$ is given as
\begin{eqnarray}
\mathbf{r} &=& a(t)\mathbf{q}+b(t)\mathbf{d}
\end{eqnarray}
where $\mathbf{q}$ is the Lagrangian coordinate
(\emph{i.e}.~initial comoving coordinate with displacement),
$\mathbf{d}$ is the displacement from $\mathbf{q}$ and $b(t)$
governs how the displacement increases in time (or, equivalently,
how the density perturbation grows). From this, we have a
deformation tensor
\begin{eqnarray}
\mathcal{D}_{ij} &=& \frac{\partial r_{i}}{\partial q_{j}}\ =\
a(t)\delta_{ij}+b(t)\frac{\partial d_{i}}{\partial q_{j}},
\end{eqnarray}
which is just the Jacobian of the coordinate transformation,
providing information about the change of the size of a given
volume element (centred on the particle). When the deformation is
small, as in the case of the early times when we set up the
initial condition, we have
\begin{eqnarray}\label{eq:Jacobian}
\det\mathcal{D} &\doteq&
a^{3}(t)\left[1+\frac{b}{a}\nabla_\mathbf{q}\cdot\mathbf{d}\right]
\end{eqnarray}
where we have neglected higher-order terms in $\mathbf{d,}$ and $\nabla _{%
\mathbf{q}}$ is the spatial derivative with respect to the Lagrangian
coordinate $\mathbf{q}$. Note that the term in brackets is the fractional
change of the volume element due to the collapse of matter.

In a given volume element, because we have two matter species now and their
density contrasts grow at different rates because of the different
scalar-field coupling, it makes sense to have two different $^{\prime
}b(t)^{\prime }$ functions: $b(t)$ for baryons and $\tilde{b}(t)$ for dark
matter. Thus, we end up with
\begin{eqnarray}\label{eq:Zel_r_B}
\mathbf{r}_{\mathrm{B}} &=& a\mathbf{q}+b\mathbf{d},\\
\label{eq:Zel_r_D}\mathbf{r}_{\mathrm{D}} &=&
a\mathbf{q}+\tilde{b}\mathbf{d},
\end{eqnarray}
in which $\mathbf{r}_{\mathrm{B}}$, $\mathbf{r}_{\mathrm{D}}$
denote the actual positions of a baryonic particle and a
dark-matter particle residing in the volume element. We now want
to solve for $b$ and $\tilde{b}$.

As we will work with constant dark-matter mass, the mass conservation is
just as simple as in $\Lambda $CDM, and, assuming no particles escape or
enter our volume element, we have
\begin{eqnarray}
\bar{\rho}_{\mathrm{B}}(t)a^{3}d^{3}\mathbf{q} &=&
\rho_{\mathrm{B}}(t,\mathbf{r})d^{3}\mathbf{r},\\
\bar{\rho}_{\mathrm{D}}(t)a^{3}d^{3}\mathbf{q} &=&
\rho_{\mathrm{D}}(t,\mathbf{r})d^{3}\mathbf{r},
\end{eqnarray}
where $\bar{\rho}$ is the average density. Now, from Eq.~(\ref{eq:Jacobian}%
), it follows directly that
\begin{eqnarray}
\rho_{\mathrm{B}}(t,\mathbf{r}) &=&
\frac{\bar{\rho}_{\mathrm{B}}(t)a^{3}}{\det\mathcal{D}}\ \doteq\
\bar{\rho}_{\mathrm{B}}(t)\left[1-\frac{b}{a}\nabla_\mathbf{q}\cdot\mathbf{d}\right],\\
\rho_{\mathrm{D}}(t,\mathbf{r}) &=&
\frac{\bar{\rho}_{\mathrm{D}}(t)a^{3}}{\det\mathcal{D}}\ \doteq\
\bar{\rho}_{\mathrm{D}}(t)\left[1-\frac{\tilde{b}}{a}\nabla_\mathbf{q}\cdot\mathbf{d}\right]
\end{eqnarray}
and so the density contrasts can be expressed in terms of $b$ and
$\tilde{b}$ as
\begin{eqnarray}\label{eq:Zel_delta_B}
\delta_{\mathrm{B}} &=&
-\frac{b}{a}\nabla_{\mathbf{q}}\cdot\mathbf{d}\ \equiv\ -D_{+}\nabla_{\mathbf{q}}\cdot\mathbf{d},\\
\label{eq:Zel_delta_D}\delta_{\mathrm{D}} &=&
-\frac{\tilde{b}}{a}\nabla_{\mathbf{q}}\cdot\mathbf{d}\ \equiv\
-\tilde{D}_{+}\nabla_{\mathbf{q}}\cdot\mathbf{d},
\end{eqnarray}
where $D_{+},\tilde{D}_{+}$ will be identified as the linear
growth factors for baryons and dark matter below.

Now consider the force laws obeyed by baryons and dark matter. For baryons
this is very simple. In the Newtonian limit:
\begin{eqnarray}\label{eq:Zel_B_forcelaw}
\ddot{\mathbf{r}}_{\mathrm{B}} &=& -\nabla_{\mathbf{r}}\phi
\end{eqnarray}
where $\nabla _{\mathbf{r}}$ is the spatial derivative with respect to $%
\mathbf{r,}$ and we have
\begin{eqnarray}
\nabla_{\mathbf{q}} &\doteq& a\nabla_{\mathbf{r}}
\end{eqnarray}
in the limit of small $\mathbf{d}$. Here, $\phi $ is the
gravitational potential given by
\begin{eqnarray}\label{eq:Zel_Pois}
\nabla_{\mathbf{r}}^{2}\phi &\doteq& 4\pi G\left[\rho_{\mathrm{B}}
+e^{\gamma\varphi}\rho_{\mathrm{D}}-2V(\varphi)\right],
\end{eqnarray}
where we have neglected the radiation, which is negligible at
later times, and the kinetic energy of the scalar field $\varphi
$, which is always small. Meanwhile, the force law for dark-matter
particles has a contribution from the scalar field coupling:
\begin{eqnarray}\label{eq:Zel_DM_forcelaw}
\ddot{\mathbf{r}}_{\mathrm{D}} &=& -\nabla_{\mathbf{r}}\phi +
\gamma\nabla_{\mathbf{r}}\varphi
\end{eqnarray}
where in case that the scalar field potential $V(\varphi )$ is
flat (as in our case) and the perturbation to the scalar field
$\varphi $ is small (for redshift $z\gtrsim 50$), the scalar field
equation of motion is approximately
\begin{eqnarray}\label{eq:Zel_scaleom}
\nabla_{\mathbf{r}}^{2}\varphi &\doteq& 8\gamma\pi
Ge^{\gamma\varphi}\left(\rho_{\mathrm{D}}-\bar{\rho}_{\mathrm{D}}\right).
\end{eqnarray}

Writing $\nabla =\nabla _{\mathbf{r}}$ from here on, and neglecting the
perturbation in $\varphi $ and $V(\varphi )$, from Eqs.~(\ref{eq:Zel_delta_B}%
, \ref{eq:Zel_Pois}) we have
\begin{eqnarray}
\nabla^{2}\phi &=& 4\pi G\left[\bar{\rho}_{\mathrm{B}} +
e^{\gamma\varphi}\bar{\rho}_{\mathrm{D}}-2\bar{V}(\varphi)\right]\nonumber\\
&& - 4\pi G\left(\bar{\rho}_{\mathrm{B}}D_{+} +
e^{\gamma\varphi}\bar{\rho}_{\mathrm{D}}\tilde{D}_{+}\right)a\nabla\cdot\mathbf{d}\nonumber\\
&=& -3\frac{\ddot{a}}{a} - 4\pi
G\left(\bar{\rho}_{\mathrm{B}}D_{+} +
e^{\gamma\varphi}\bar{\rho}_{\mathrm{D}}\tilde{D}_{+}\right)a\nabla\cdot\mathbf{d}\nonumber
\end{eqnarray}
where we have used the Raychaudhuri equation. This can be
integrated once to obtain
\begin{eqnarray}\label{eq:Zel_grad_phi}
\nabla\phi &=& -\frac{\ddot{a}}{a}\mathbf{r} - 4\pi
G\left(\bar{\rho}_{\mathrm{B}}D_{+} +
e^{\gamma\varphi}\bar{\rho}_{\mathrm{D}}\tilde{D}_{+}\right)a\mathbf{d}.
\end{eqnarray}
Eqs.~(\ref{eq:Zel_r_B}, \ref{eq:Zel_B_forcelaw},
\ref{eq:Zel_grad_phi}) then give, after some algebra,
\begin{eqnarray}\label{eq:Zel_eq_D_B}
\ddot{D}_{+} + 2\frac{\dot{a}}{a}\dot{D}_{+} - 4\pi
G\left[\bar{\rho}_{\mathrm{B}}D_{+} +
e^{\gamma\varphi}\bar{\rho}_{\mathrm{D}}\tilde{D}_{+}\right] &=&
0.\ \ \ \
\end{eqnarray}
This is the equation for $b$.

The idea of deriving the equation for $\tilde{b}$ is quite similar, but now
we need to take into account the fifth force. Here, Eq.~(\ref{eq:Zel_scaleom}%
) can be re-expressed, using Eq.~(\ref{eq:Zel_delta_D}), as
\begin{eqnarray}
\nabla^{2}\varphi &=&  8\pi Ga\gamma
e^{\gamma\varphi}\bar{\rho}_{\mathrm{D}}\tilde{D}_{+}\nabla\cdot\mathbf{d}\nonumber
\end{eqnarray}
which can be integrated once to obtain
\begin{eqnarray}\label{eq:Zel_grad_varphi}
\nabla\varphi &=&  8\pi Ga\gamma
e^{\gamma\varphi}\bar{\rho}_{\mathrm{D}}\tilde{D}_{+}\mathbf{d}.
\end{eqnarray}
Then, Eqs.~(\ref{eq:Zel_r_D}, \ref{eq:Zel_DM_forcelaw}, \ref{eq:Zel_grad_phi}%
, \ref{eq:Zel_grad_varphi}) lead, again after some algebra, to
\begin{eqnarray}\label{eq:Zel_eq_D_D}
\ddot{\tilde{D}}_{+} + 2\frac{\dot{a}}{a}\dot{\tilde{D}}_{+} -
4\pi G\left[\bar{\rho}_{\mathrm{B}}D_{+} + \beta
e^{\gamma\varphi}\bar{\rho}_{\mathrm{D}}\tilde{D}_{+}\right] &=&
0,\ \ \ \
\end{eqnarray}
where we have defined $\beta \equiv 1+2\gamma ^{2}$ to encode the
effects of the fifth force.

Eqs.~(\ref{eq:Zel_eq_D_B}, \ref{eq:Zel_eq_D_D}) indicate that $D_{+},\tilde{D%
}_{+}$ are the linear growth factors of baryons and dark matter. Indeed, if
the coupling constant $\gamma =0$, then we find that $D_{+}=\tilde{D}_{+}$
and
\begin{eqnarray}
\ddot{D}_{+} + 2\frac{\dot{a}}{a}\dot{D}_{+} - 4\pi
G\bar{\rho}_{m}D_{+} &=& 0,\ \ \ \
\end{eqnarray}
where
$\bar{\rho}_{m}=\bar{\rho}_{\mathrm{B}}+\bar{\rho}_{\mathrm{D}}$
is the total matter density. This is the familiar equation of
linear growth equation in this approximation.

In this discussion we have made several approximations. For example, the
kinetic energy of scalar field, which is always subdominant, is neglected;
the perturbation of the scalar field potential energy is also neglected
since the scalar field perturbation is small, especially at the time we set
up the initial condition; the spatial dependence of $\tilde{D}_{+}$ is
neglected implicitly, because we only consider small scales where the fifth
force simply rescales the gravitational constant which governs structure
growth.

Eqs.~(\ref{eq:Zel_delta_B}, \ref{eq:Zel_delta_D}) are the starting point of
the numerical codes which generate initial conditions for N-body
simulations, such as \texttt{GRAFIC2}. Given the matter power spectrum at
the initial time $t_{i}$, the code produces density fluctuation field $%
\delta $ as a Gaussian random field and works out the displacement field $%
\mathbf{d}$, or actually $D_{+}\mathbf{d}$, because
\begin{eqnarray}\label{eq:Zel_displace}
\mathbf{r} &=& a\mathbf{q}+b\mathbf{d}\ =\
a\left(\mathbf{q}+D_{+}\mathbf{d}\right)\ \equiv\ a\mathbf{x}.
\end{eqnarray}
The initial peculiar velocity of the particle is then
\begin{eqnarray}\label{eq:Zel_vel}
\mathbf{v} &=& a\dot{\mathbf{x}}\ =\ a\dot{D}_{+}\mathbf{d}\
\equiv\ f\frac{d\ln a}{d\tau}D_{+}\mathbf{d}
\end{eqnarray}
where $\tau$ is the conformal time and
\begin{eqnarray}
f &\equiv& \frac{d\ln D_{+}}{d\ln a}.
\end{eqnarray}
In our model the initial displacements and velocities of particles
should be generated separately for the two different matter
species, using their
respective matter power spectrum and linear growth factor $D_{+}$ (or $%
\tilde{D}_{+}$).

\end{document}